\documentclass[manuscript,screen]{acmart}
\usepackage{lineno}
\usepackage{tabularx}
\usepackage{amsmath}
\usepackage{algorithmic}
\usepackage{subcaption}
\usepackage{graphicx}
\usepackage{textcomp}
\usepackage{multirow}
\usepackage{adjustbox}
\usepackage{anyfontsize}
\usepackage{makecell}
\usepackage{array}
\usepackage{colortbl}
\usepackage{caption}
\usepackage{url}

\modulolinenumbers[5]
\AtBeginDocument{%
  \providecommand\BibTeX{{%
    \normalfont B\kern-0.5em{\scshape i\kern-0.25em b}\kern-0.8em\TeX}}}

\setcopyright{acmcopyright}
\copyrightyear{2018}
\acmYear{2018}
\acmDOI{XXXXXXX.XXXXXXX}

\acmConference[Conference acronym 'XX]{Make sure to enter the correct
  conference title from your rights confirmation emai}{June 03--05,
  2018}{Woodstock, NY}
\acmPrice{15.00}
\acmISBN{978-1-4503-XXXX-X/18/06}

\begin{document}

\title{A Requirement-Based Framework for Engineering Adaptive Authentication}

\author{Alzubair Hassan}
\email{alzubair.hassan@ucd.ie}
\orcid{1234-5678-9012}
\authornotemark[1]
\affiliation{%
  \institution{University College Dublin and Lero}
  \city{Dublin}
  \country{Ireland}
}

\author{Alkabashi Alnour}
\email{alkabashi.alnour@ucdconnect.ie}
\affiliation{%
  \institution{University College Dublin and Lero}
  \city{Dublin}
  \country{Ireland}
}

\author{Bashar Nuseibeh}
\affiliation{%
  \institution{Open University}
  \city{Milton Keynes}
  \country{UK}}
\email{bashar.nuseibeh@lero.ie}

\author{Liliana Pasquale}
\affiliation{%
  \institution{University College Dublin and Lero}
  \city{Dublin}
  \country{Ireland}
}
\email{liliana.pasquale@ucd.ie}

\renewcommand{\shortauthors}{Hassan et al.}

\begin{abstract}

Authentication is crucial to confirm that an individual or entity trying to perform an action is actually who or what they claim to be. In dynamic environments such as the Internet of Things (IoT), Internet of Vehicles (IoV), healthcare, and smart cities,  security risks can change depending on varying contextual factors (e.g., user attempting to authenticate, location, device type). Thus, authentication methods must adapt to mitigate changing security risks while meeting usability and performance requirements. However, existing adaptive authentication systems provide limited guidance on (a) representing contextual factors, requirements, and authentication methods (b) understanding the influence of contextual factors and authentication methods on the fulfilment of requirements, and (c) selecting effective authentication methods that reduce security risks while maximizing the satisfaction of the requirements. This paper proposes a framework for engineering adaptive authentication systems that dynamically select effective authentication methods to address changes in contextual factors and security risks. The framework leverages a contextual goal model to represent requirements and the influence of contextual factors on security risks and requirement priorities. It uses an extended feature model to represent potential authentication methods and their impacts on mitigating security risks and satisfying requirements. At runtime, when contextual factors change, the framework employs a Fuzzy Causal network encoded using the Z3 SMT solver to analyze the goal and feature models, enabling the selection of effective authentication methods. We demonstrate and evaluate our framework through its application to real-world authentication scenarios in the IoV and the healthcare domains. 

\end{abstract}



\maketitle

\section{Introduction}
Authentication is the process of verifying the identity of a person, device, or system to ensure they are authorized to access a resource or perform an action~\cite{Menezes2018}. Authentication methods typically require credentials based on something you know (e.g., passwords, PINs), something you have (e.g., smart cards, security keys, or mobile phones) or something you are (e.g.,  fingerprints, face, iris, or voice recognition). 

In dynamically changing environments such as the Internet of Things (IoT), Internet of Vehicles (IoV), healthcare, and smart cities, security risks and the requirements (e.g., usability and performance) priorities can change
depending on contextual factors (e.g., user attempting to authenticate, location, device type)~\cite{hassan2021engineer}. Also, certain factors (e.g., night time) can hinder the feasibility of some authentication methods (e.g., face recognition). Traditional authentication methods relying on static credentials may not protect these systems from changing security risks or may not satisfy their requirements sufficiently.
Thus, an adaptive authentication approach should adapt the authentication methods to mitigate changing security risks while meeting the system requirements. 
For example, suppose a user attempts to access data (including personal and medical information) from different locations and prefers an authentication method that does not require them to enter a complex password.
If the user attempts to access medical information from home, the security risk is low, and a biometric-based authentication method can be adopted in this situation. However, the security risk increases if the user attempts to access their medical records using an open public transport network. Two-factor authentication is more appropriate to mitigate these increased security risks. 


As adaptive authentication systems must respond to changes in contextual factors, they require context models\textcolor{blue}{~\cite{bumiller2023understand,chen2025context}}. Such models should accurately represent contextual factors' impact on security risks, requirements priorities and the feasibility of authentication methods.
 Suppose relevant contextual factors and their impacts are not represented accurately. In that case, adaptive authentication systems can enable authentication methods that leave the system unprotected or do not satisfy relevant system requirements. Previous research on adaptive authentication systems (e.g., ~\cite{bumiller2023understand,arias2019survey,bakar2013adaptive}) offers limited guidance on (a) representing contextual factors, requirements, and authentication methods (b)
understanding the influence of contextual factors and authentication methods on the level of security risks and the satisfaction of requirements, and (c) selecting
 authentication methods that reduce security risks while satisfying the requirements.

To address these limitations, this paper aims to model contextual factors, system requirements, security risks and authentication methods to support adaptive selection of authentication methods in response to changing risk and requirements priorities at runtime.
We propose a framework for engineering
adaptive authentication systems that dynamically select effective authentication methods that minimize security risks and maximize requirements satisfaction. In our previous work~\cite{hassan2021engineer}, we reviewed previous research to characterize the adaptive authentication problem and support the development of an adaptive authentication system. 
 In this paper, we extend our previous work by detailing how contextual factors, authentication methods, and requirements should be represented, and we provide a technique to automatically select authentication methods. We use a contextual goal model~\cite{ali2010goal} to represent the requirements and the impact of contextual factors on security risks, priority of requirements, and feasibility of authentication methods. We use an extended feature model~\cite{kang1990feature} to represent the features of different authentication methods and their impact on the satisfaction of the requirements. We use a Fuzzy Causal Network (FCN)~\cite{FCN} encoded using Z3~\cite{z3}, an efficient SMT solver, to reason about the impact of contextual factors on security risks and requirements priorities and select an effective authentication method that can be applied in the given context.

 We evaluate our framework against a baseline, static authentication approach that selects authentication methods based on contextual factors and requirements.  We have applied our framework to two authentication scenarios in the IoV and healthcare domains. We demonstrate that our framework always chooses an effective authentication method at runtime. For example, our framework chooses a strong authentication method (e.g., two-factor authentication) when the security risk and the priority of security requirements are high. Conversely, it chooses a usable authentication method (e.g., face recognition) when the usability requirement has a higher priority than other requirements. Additionally, it selects an alternative authentication method (e.g., fingerprint) when contextual factors (e.g., nighttime) render the current authentication method (e.g., iris and face recognition) ineffective. We also demonstrate that our approach has a small computational overhead and can be used at runtime.
 Our work contributes to the self-adaptive systems community by demonstrating how the integration of goal and feature models can support context-aware, requirement-driven adaptation at runtime. Our framework informs authentication decisions in dynamic environments, bridging the gap between design-time modeling and runtime decision-making. Furthermore, it positions security—specifically, adaptive authentication—as a novel and critical application domain for self-adaptive systems, advancing its real-world applicability.

\section{Authentication Scenarios in IoV}
\label{sec:Scenarios}
\begin{figure*}[h]
 	\centering
 	\begin{subfigure}[b]{0.32\linewidth}
 		\includegraphics[width=\linewidth]{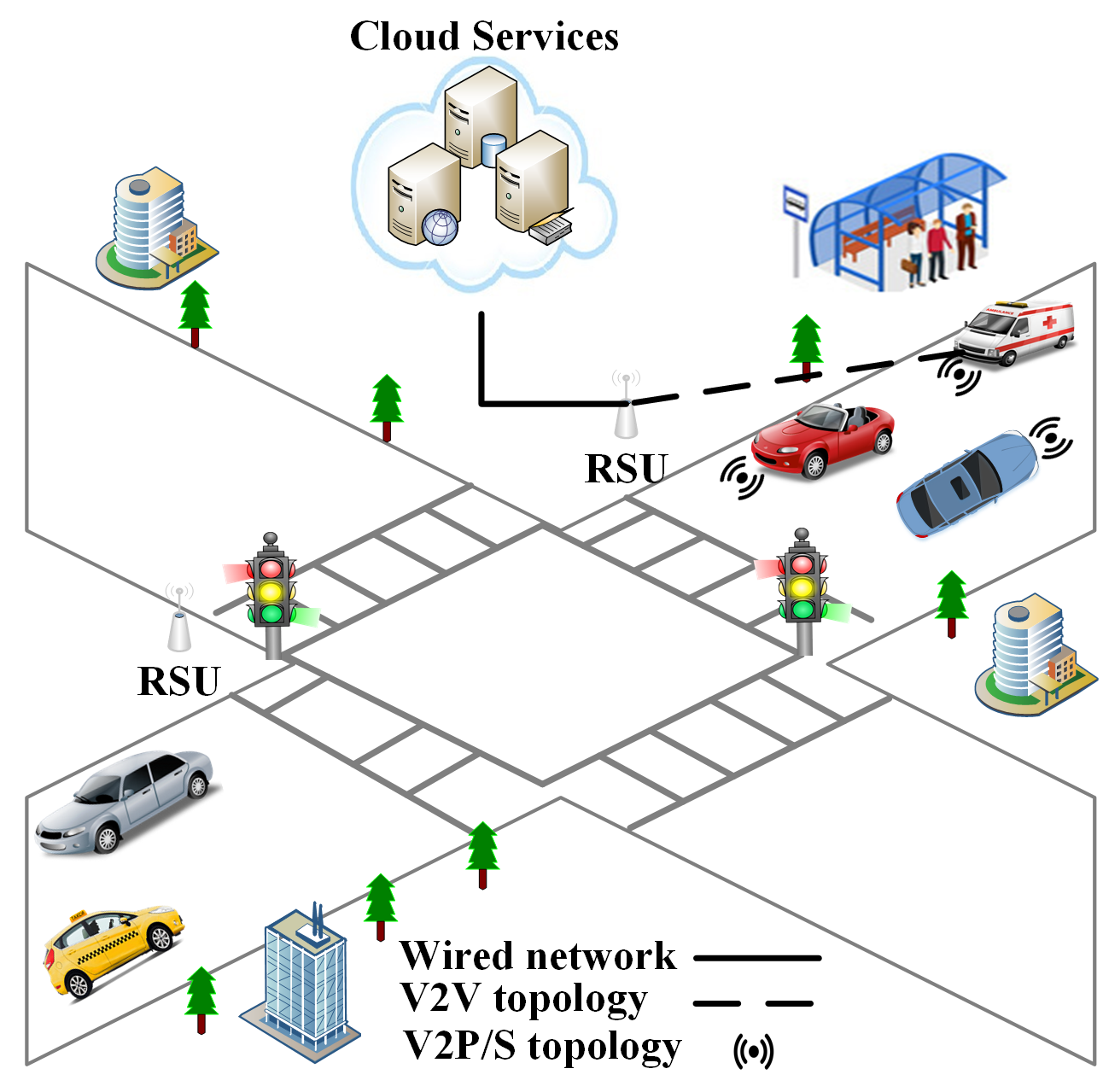}
 		\caption{Secure Data Exchange }
 		\label{fig-3a}
 	\end{subfigure}
 	\begin{subfigure}[b]{0.32\linewidth}
 		\includegraphics[width=\linewidth]{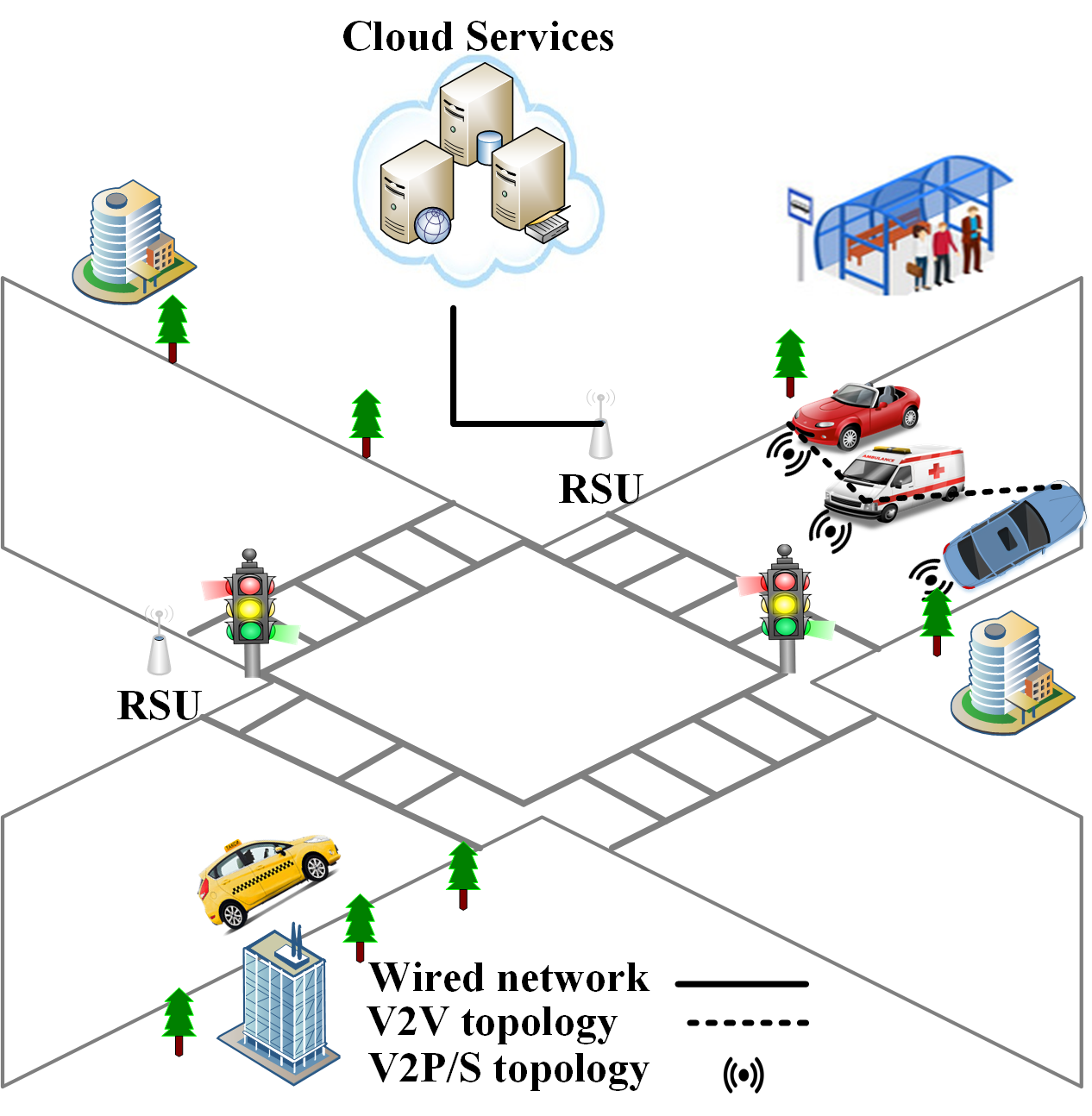}
 		\caption{Ambulance Overtaking Another Vehicle }
 		\label{fig-3b}
 	\end{subfigure}
 \begin{subfigure}[b]{0.32\linewidth}
 	\includegraphics[width=\linewidth]{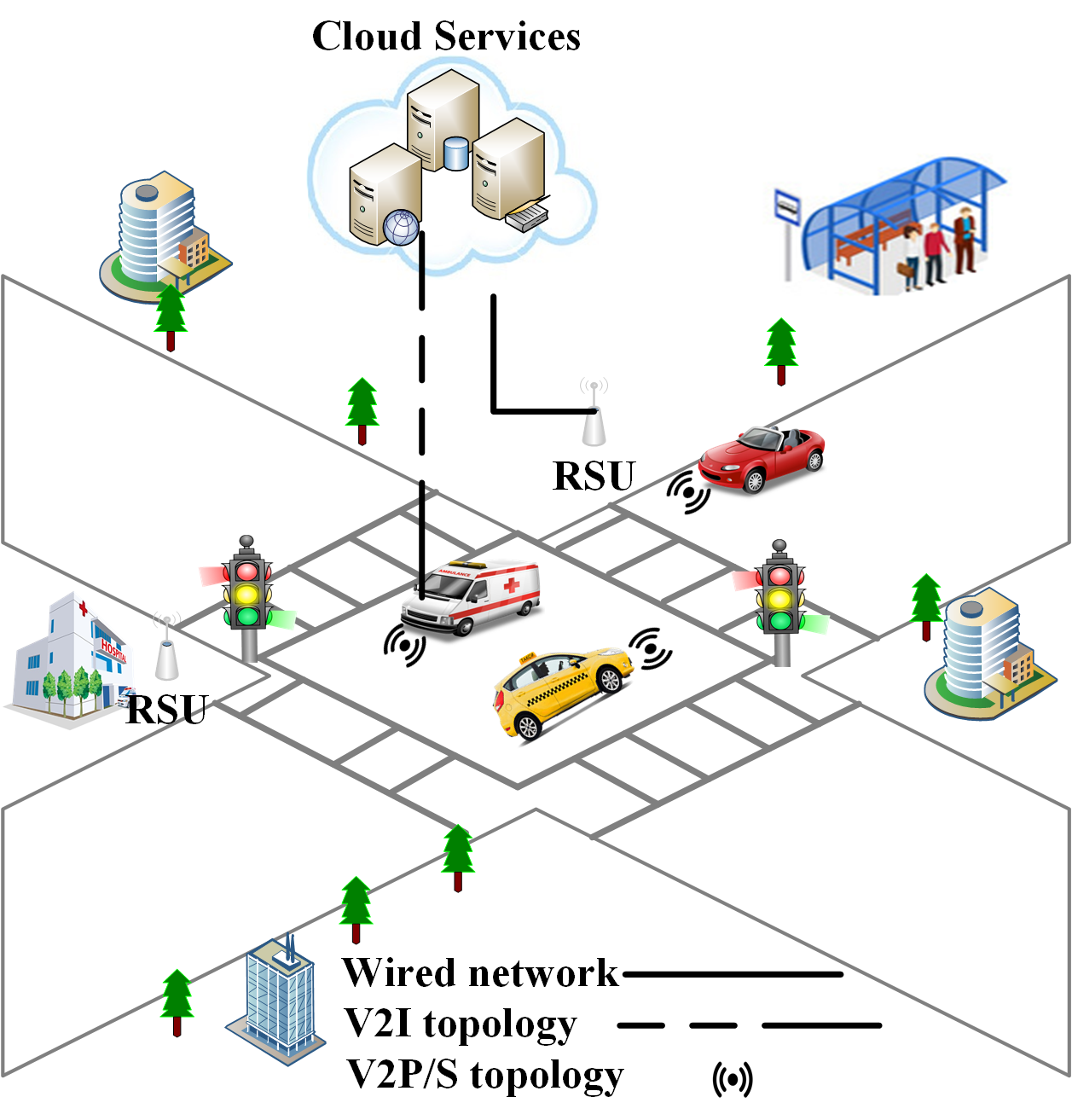}
 	\caption{Accessing Patient Data at a Junction}
 	\label{fig-3c}
 \end{subfigure}
 	\caption{IoV Adaptive Authentication Scenarios.}
 	\label{fig-3}
 \end{figure*}

Our motivating example is informed by potential attacks in different IoV network topologies and applications/services~\cite{sharma2019survey,ali2019authentication} (see Table~\ref{tab-1}). The IoV network is a heterogeneous vehicular network combining inter-vehicle and intra-vehicle networks and vehicular mobile Internet~\cite{kaiwartya2016}. As shown in Figure~\ref{fig-3}, an IoV network can include users (e.g., drivers, passengers, and pedestrians), vehicles (e.g., cars, buses) and devices (e.g., mobile devices, roadside units or RSUs). Users, vehicles and devices can capture and share sensitive information, such as citizens' personal and road traffic information. Vehicles and devices can communicate using different topologies, such as Vehicle-to-Vehicle (V2V), Vehicle-to-Roadside units (V2R), and Vehicle-to-Infrastructure (V2I) via cellular networks.

 \begin{table}[htbp]
 	\caption{Examples of attacks in different IoV network topologies and applications/services~\cite{sharma2019survey}.}
 		\begin{tabular}{|p{2cm}|p{4cm}|p{5.5cm}|}
 			\hline
 			\textbf{IoV topology} & \textbf{\textit{ Attacks/threats }} & \textbf{\textit{Application/services}}\\
			\hline
 			V2V & V2V timing attack, replay attack, impersonation &  SOS services, post-crash warning, wrong way driver warning \\
			\hline
 	V2R &  DoS, impersonation & intersection collision avoidance, in-vehicle signage \\
 		\hline
 	V2I &  timing attack & intersection collision warning, public safety related applications, SOS services, post-crash warning \\
 		\hline
  			\end{tabular}
	\label{tab-1}

 \end{table}
 
The ambulance requires real-time road traffic information to reach the hospital quickly. To achieve this aim, it communicates with the nearest RSUs using a Vehicle-to-Roadside Units (V2R) communication topology (see Figure~\ref{fig-3a}). However, nearby vehicles may attempt to impersonate the ambulance to obtain road traffic information illegitimately. In this scenario, ensuring the confidentiality of traffic data and the authenticity of the communicating parties (the ambulance and the RSU) takes precedence over usability and performance considerations. To mitigate the risk of impersonation attacks, the ambulance and RSU should employ secure methods such as certificate-based or signcryption-based authentication before exchanging information.

In another situation, the ambulance attempts to overtake a red car (see Figure~\ref{fig-3b}). This requires exchanging distance information between the ambulance and nearby vehicles (red and blue cars) using a Vehicle-to-Vehicle (V2V) communication topology. Ensuring the integrity of this distance information is critical to preventing collisions. Additionally, the exchange must occur quickly so the ambulance can overtake the red car without delay. Here, performance requirements, such as minimizing authentication time, are prioritized over security and usability concerns. Certificate-based authentication is unsuitable in this case, as it may involve lengthy identity verification through a remote server. Instead, vehicles can use alternative methods, such as verifying each other with car plates and driver’s licenses, which offer faster authentication while reducing the risk of impersonation attacks.

The ambulance driver is approaching a junction and accessing patient information through a Vehicle-to-Infrastructure (V2I) connection via cellular networks (see Figure~\ref{fig-3c}). Since the information is sensitive, ensuring its confidentiality is crucial. At the same time, usability is also a key consideration, as the authentication process should not distract the driver, who needs to remain focused on navigating the junction. A biometric-based authentication method, such as face or iris recognition, would be ideal in this situation, as it does not require the driver to perform an action (e.g., entering a password or swiping a card). 

Besides impersonation attacks, IoV networks can also be subject to timing attacks and Denial-of-Service (DoS) attacks. Timing attacks exploit variations in the execution time of security operations (e.g., authentication or cryptographic verification) to infer sensitive information, which may occur in Vehicle-to-Vehicle (V2V) communication scenarios. In contrast, a Denial-of-Service (DoS) attack overwhelms a system with excessive requests, rendering it unavailable to legitimate users, a threat particularly relevant in Vehicle-to-Roadside Unit (V2R) settings. When the risk of such attacks is high,  digital signatures or signcryption can introduce variability and complexity, making it harder for attackers to exploit timing differences~\cite{brumley2005remote}. However, cryptographic authentication alone does not inherently prevent timing attacks or DoS attacks~\cite{hassan2022secure}. Timing attacks are more effectively mitigated by using constant-time cryptographic implementations, while DoS attacks require complementary mechanisms such as rate limiting, request filtering, or distributed DoS mitigation strategies~\cite{mirkovic2004taxonomy}. Adaptive authentication can contribute by selecting more robust mechanisms under high-risk conditions, but it must be combined with additional protections for comprehensive defense.

Contextual factors, such as location, network topology, the sensitivity of the accessed information, and the proximity of other vehicles, can influence the security risks and the priority of requirements in adaptive authentication (e.g., security, usability, and performance). These requirements may even conflict with one another. For instance, a certificate-based authentication can negatively impact performance, or a highly complex password can undermine usability. Certain factors, such as low lighting, might also render some authentication methods (e.g., face recognition) ineffective. Additionally, it is difficult to precisely quantify the impact of a given authentication method on satisfying these requirements. Finally, as users actively engage in authentication, their preferences and privacy needs must be considered when choosing an authentication method.

Therefore, it is necessary to develop an approach for systematically representing contextual factors, their impact on authentication requirements, and the feasibility of authentication methods. This approach should effectively capture the dynamic interplay between security risks, user needs, and authentication mechanisms in response to changing contextual conditions. 

\begin{figure}[h]
\centering
		\includegraphics[width=0.4\columnwidth] {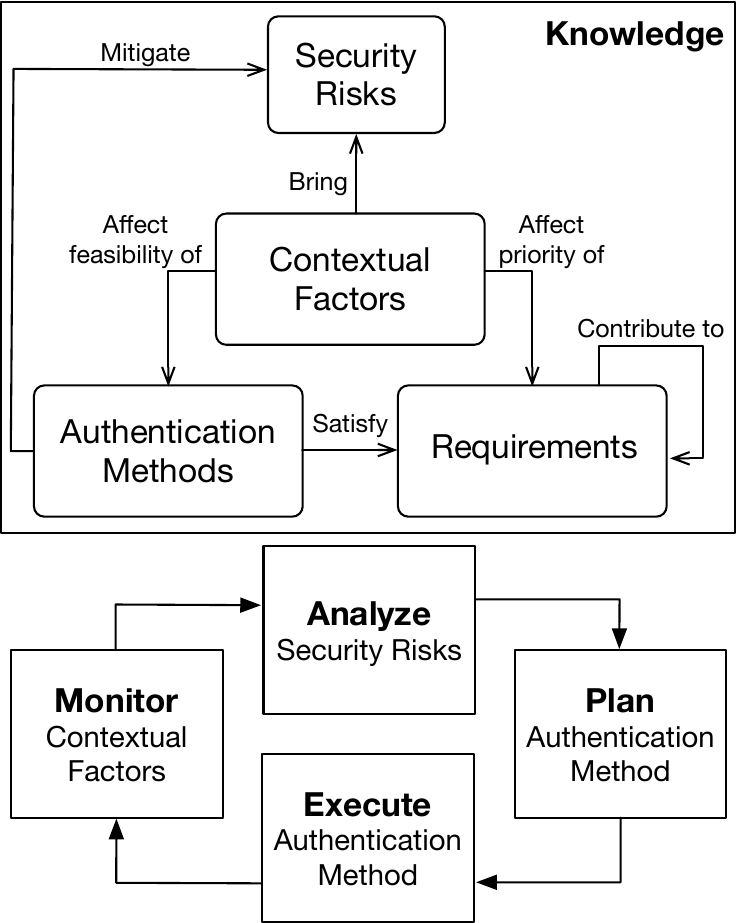}
	\caption{Adaptive Authentication System}
	\label{fig:Framework}
\end{figure}

\section{Adaptive Authentication Framework}
\label{sec:framework}

We propose a framework to build an adaptive authentication system based on the MAPE-K (Monitor, Analyze, Plan, Execute, Knowledge) loop~\cite{kephart2003vision}~\cite{Arcaini.SEAMS.2015} shown in Figure~\ref{fig:Framework}. The Knowledge of the MAPE-K loop is based on the representation of the requirements, contextual factors and authentication methods that can be selected at runtime~\cite{hassan2021engineer}. Contextual factors can bring security risks and affect the priority of the requirements. They can also make specific authentication methods infeasible. Authentication methods, instead, can mitigate security risks and help satisfy the requirements. The adaptive authentication system should monitor contextual factors and analyze security risks during monitoring and analysis. During planning, an effective authentication method should be identified that a) is feasible, b) minimizes security risks, and c) maximizes the satisfaction of the requirements considering their trade-offs. During execution, the adaptive authentication system should enforce the selected authentication method in the adapted system. We use a MAPE-K loop architecture to support the separation of concerns between context monitoring, security risk analysis, decision-making, and the enforcement of authentication methods.

\begin{figure}[htpb]
\centering
		\includegraphics[width=0.7\columnwidth] {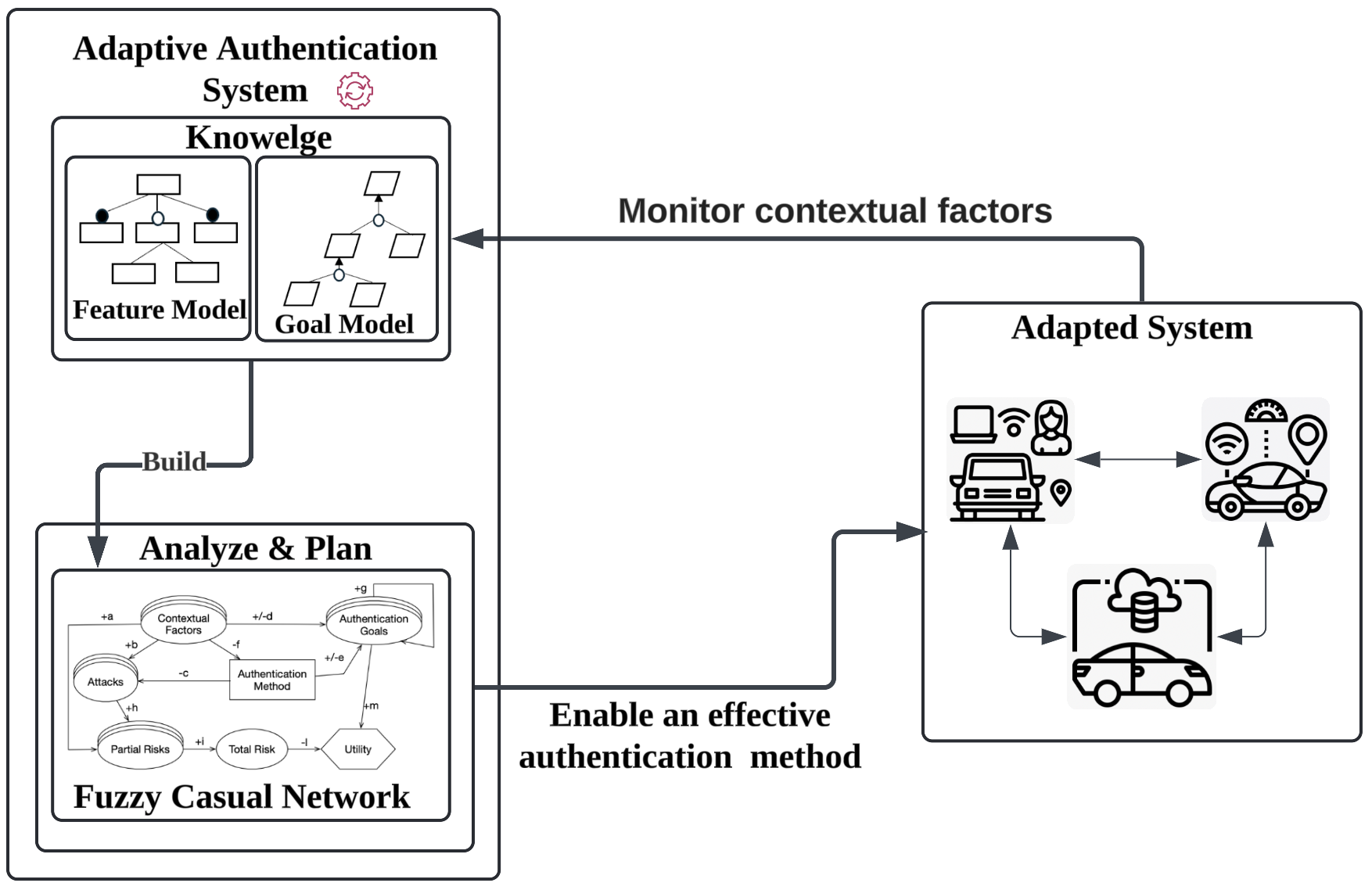}
	\caption{Our approach for engineering adaptive authentication }
	\label{fig:our}
\end{figure}

This paper does not address how contextual factors can be monitored or how adaptive authentication actions can be implemented in the system. Instead, it focuses on how to represent contextual factors, requirements, and authentication methods and their interdependencies, and on how to compute the level of security risk based on contextual factors and select an effective authentication method. 
Figure \ref{fig:our} provides an overview of our approach for engineering adaptive authentication systems.  To achieve this aim, we use a contextual goal model~\cite{ali2010goal} to represent the requirements and the impact of contextual factors on the requirements priorities. We use an extended feature model~\cite{kang1990feature} to represent authentication methods and their impact on the satisfaction of the requirements. Also, we explicitly represent how the contextual factors affect the feasibility of the authentication methods.  To select an effective authentication method, we leverage a Fuzzy Causal Network (FCN)~\cite{FCN}, which has been used in previous work on adaptive security~\cite{Salehie.RE.2012} to analyze the consequences of context and asset changes on
security risks and perform impact analyses to select security controls. To reduce the time to compute an effective authentication method, we encode the FCN in Z3, an efficient SMT solver developed by Microsoft Research~\cite{z3}. It is primarily used for formal verification, constraint solving, and automated reasoning in various domains, including software verification, security analysis, and artificial intelligence.

\subsection{Knowledge Representation}

\subsubsection{Goal Model}
\label{sec:goalModel}

\begin{figure}[htbp]
	\centering
		\includegraphics[width=0.8\columnwidth] {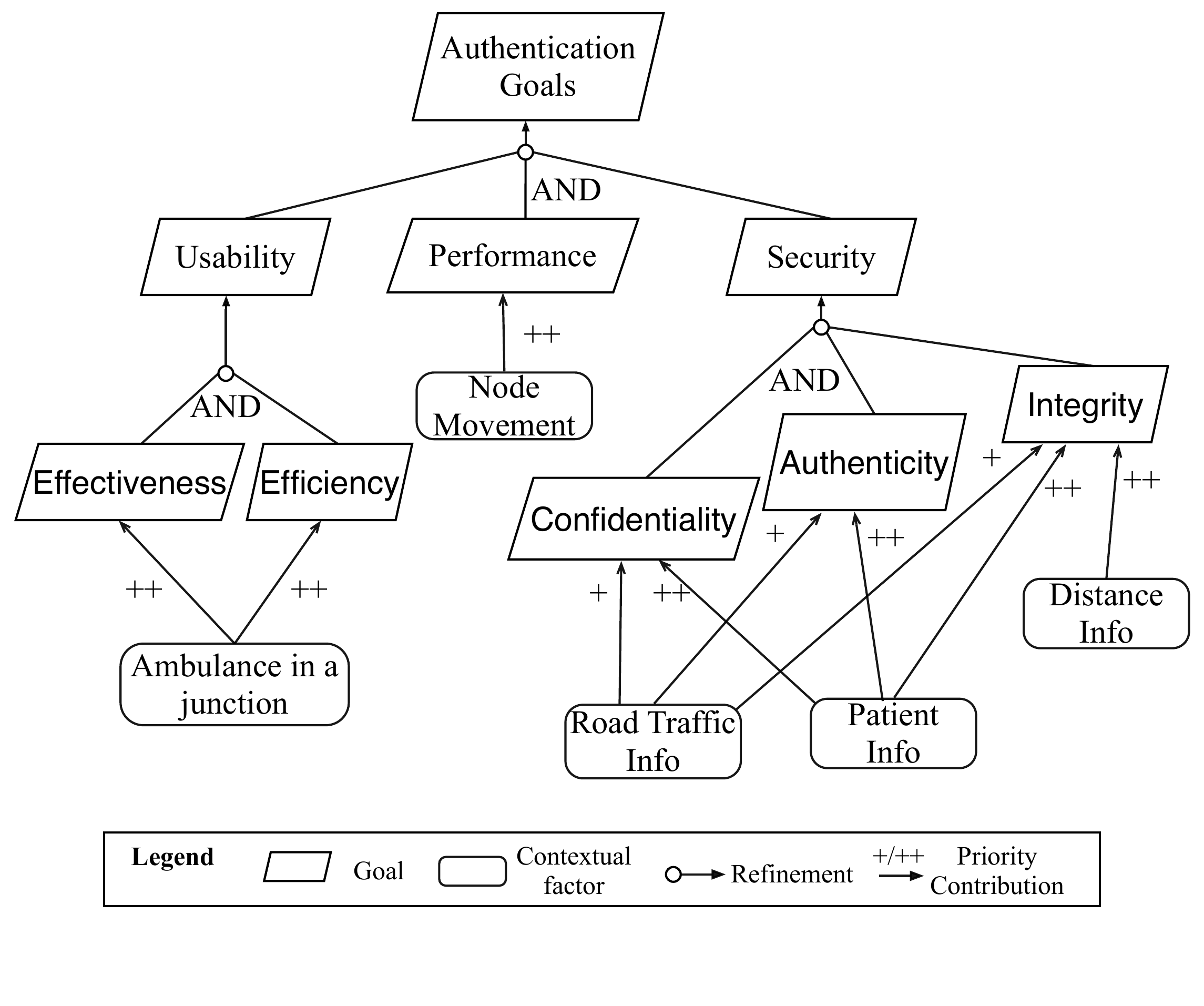}
	\caption{Requirements and Contextual Factors}
	\label{fig:goals}
	
\end{figure}

Goal models~\cite{van2009requirements} allow the representation of system requirements.
Root goals are decomposed into more concrete, realizable sub-goals, leading to identifying requirements (leaf goals). The AND-refinement means that all sub-goals (if necessary) contribute to the satisfaction of the upper-level goal.  
 Figure~\ref{fig:goals} depicts the goal model associated with the scenarios described in Section~\ref{sec:Scenarios}. An authentication system typically aims to satisfy{\it security} goals, such as {\it Confidentiality}, {\it Authenticity}, and {\it Integrity}~\cite{Margulies.2015}. 
{\it Confidentiality} aims to keep sensitive information and authentication credentials (e.g., passwords, biometric data, cryptographic keys) private and protect them from unauthorized access. {\it Integrity} aims to maintain authentication data and communications unaltered and tamper-proof. {\it Authenticity} ensures that a user, device, or system authenticating is who or what they claim to be. Security requirements are operationalized by authentication methods.
{\it Effectiveness} and {\it Efficiency} decompose the  {\it Usability} goal.{\it Effectiveness} aims to minimize the user error rate associated with an authentication method. This can be influenced by factors such as credential memorability (e.g., a difficult-to-remember password may reduce effectiveness) and environmental factors (e.g., noise, lighting conditions, or temperature) that may hinder the user's ability to authenticate correctly~\cite{wojtowicz2016model}.  {\it Efficiency } refers to how smoothly and quickly a user can complete the authentication process with minimal effort. It also involves reducing unnecessary steps, simplifying interactions, and minimizing the required credentials (the number of tries needed before successful authentication)~\cite{riva2012progressive,jorquera2018improving}.  Note that {\it Effectiveness} and {\it Efficiency} requirements decompose the {\it Usability} goal because they refer to the user ability to complete the authentication without errors and quickly. While, the {\it Performance} requirement aims at minimizing the time needed by the system to validate the credentials entered by the user during authentication. 


\

The priority of a leaf goal (or requirement) is influenced by contextual factors. We incorporate contextual factors within the goal model using contextual goal models~\cite{ali2010goal}. These have been used in adaptive systems to facilitate goal-driven decision-making in dynamic environments~\cite{chen2025context}. In our framework, we use a contextual goal model to represent the impact of contextual factors on requirements priorities.

In this paper, we quantify the impact with a value in $[0,1]$. 
Values greater than $0.8$ are interpreted as indicating a very strong positive impact (++). Values in $(0.6, 0.8]$ represent a positive impact (+). Values in $(0.4, 0.6]$ indicate a moderate impact just above the neutral midpoint (0.5). Values less than equal $0.2$ indicate a very strong negative impact (--) , while values in $(0.2,04]$ indicate a negative impact (-).
The chosen intervals are intended as a transparent and operational mapping from qualitative labels to a normalized $[0,1]$ scale.

 For example, when the ambulance is at a junction, the priority of the {\it Effectiveness} and {\it Efficiency} requirements is very high ($++$), i.e. $> 0.8$, because the driver needs to perform authentication without distraction when they need to focus on crossing the junction~\cite{kalamandeen2010ensemble}. Authenticating when cars are moving ({\it Node Movement}) affects the priority of the {\it Performance requirement} ($> 0.8$), since authentication must occur quickly~\cite{kalamandeen2010ensemble}. When, road traffic information is transmitted ({\it Road Traffic Info}) the priority of {\it Confidentiality},  {\it Integrity}, and {\it Authenticity} requirements is high ($+$) (in $(0.6,0.8]$), since such information should be accessed and modified only by trusted users (e.g., ambulances). When the ambulance exchanges distance information with a nearby vehicle, the integrity goal has a very high priority because tampering with this information could cause a crash between the two vehicles.

 If more than one contextual factor affects the priority of the same goal or requirement, we aggregate their impact by considering the maximum. This ensures that if any individual factor signals a high threat, the system enforces the strongest authentication method to mitigate the risk. This choice reflects a conservative, risk-aware strategy in which the most critical contextual factor dominates the prioritization. In security-sensitive settings, a single high-impact condition is sufficient to justify elevating a goal’s priority, regardless of the presence of additional weaker factors. While this aggregation does not capture cumulative effects, it avoids overestimating priority due to compounding assumptions and keeps the model simple. Exploring alternative aggregation functions is left for future work.

 In summary, the priority of a goal is computed as the maximum of the contributions determined by the related contextual factors. This is indicated in Eq. 1, where $g$ is a leaf goal, $CF(g)$ is the set of contextual factors that are linked to a goal, and $impact(c_i, g)$ is the impact of contextual factor $c_i$ to goal $g$. $P(g)$ is the priority of $g$.

   \begin{equation}
    P(g) = \max_{1 \le i \le N} \operatorname{impact}(c_i, g),
\quad CF(g) = \{c_1, \ldots, c_N\}.
\end{equation}

The security goals and requirements we consider in this paper are non-functional and can be partially satisfied; their degree of satisfaction is quantified as a number in $[0,1]$~\cite{Glinz.2007}. The satisfaction level of a non-leaf goal is determined by taking the minimum satisfaction value among its sub-goals, as shown in Eq. 1.

\begin{equation}
 S(g) = \min \{\, S(h) \mid h \in children(g)\,\}.
\end{equation}

 We chose the minimum operator because it has been used to express the semantics of the AND operator between fuzzy variables ($\in [0,1]$)~\cite{Zadeh.1965,Baresi.2010}. For example, the satisfaction of the {\it Usability} goal is computed as the minimum between the satisfaction of the {\it Effectiveness} and {\it Efficiency} goals.

 The satisfaction of a leaf goal, instead, is computed by aggregating a) its priority (discussed in Eq. 1) and b) the contribution of the enabled authentication methods towards its satisfaction (discussed in Section~\ref{sec:Feature}).

\subsubsection{Extended Feature Model}
\label{sec:Feature}
\begin{figure*}[!h]
	\centerline{
		\includegraphics[width=1\columnwidth] {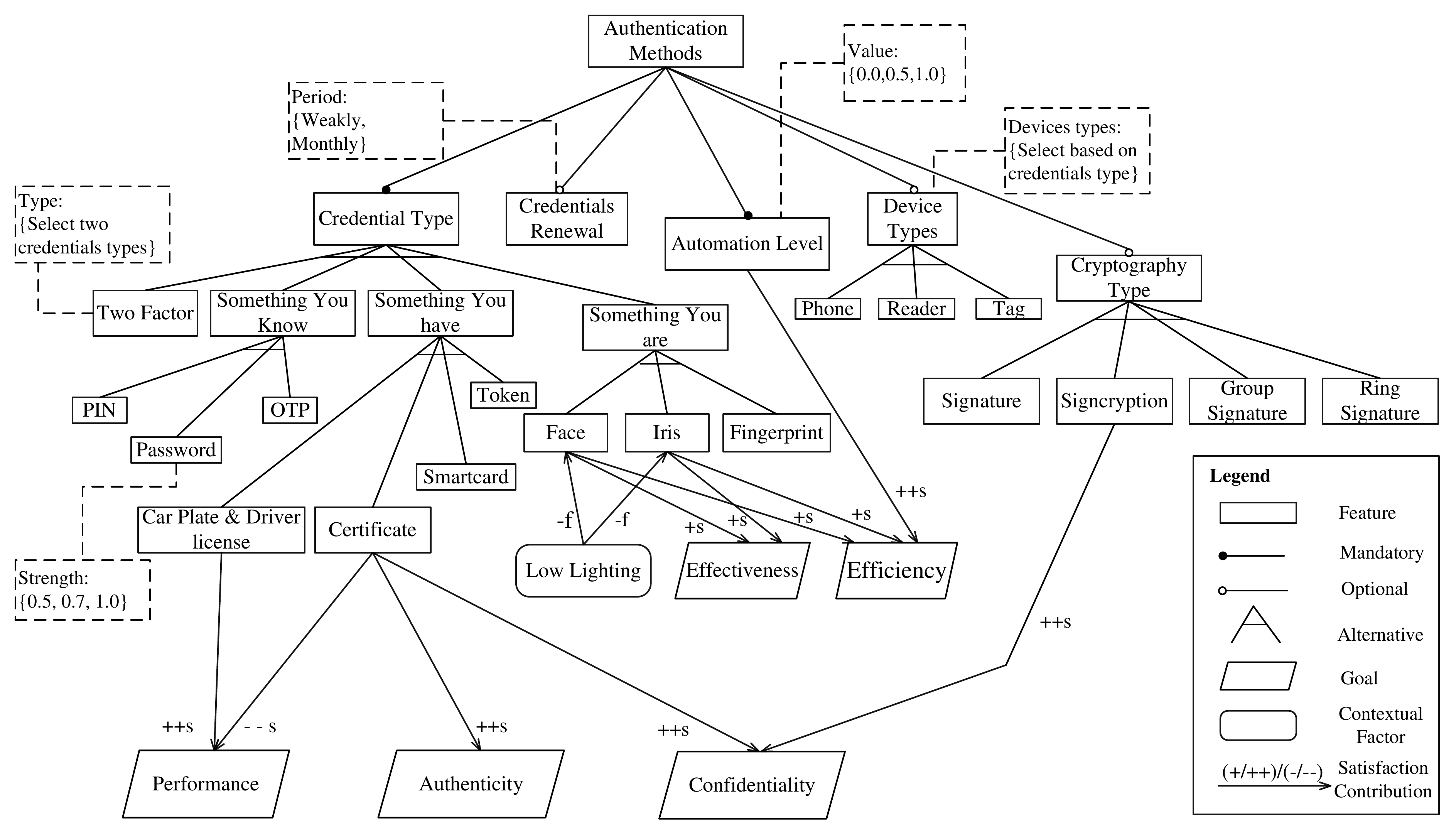}}
	\caption{Extended Feature model of authentication methods}
	\label{fig:features}
\end{figure*}

We utilize an extended feature model~\cite{kang1990feature} to represent the features that can be selected to customize an authentication method during planning. Extended feature models are widely used in software engineering and industry to capture and manage variability in systems and products. Figure~\ref{fig:features} illustrates the extended feature model, showcasing possible authentication methods. The upper section of the figure distinguishes between mandatory authentication features (e.g., credential type and automation level) and optional features (e.g., credentials renewal and device type).  {\it Credential Type} can be based on {\it Something you Know} (PIN, Password, one-time password or OTP), {\it Something you Have} (Car Plate \& Driver's License, Certificate, Smartcard or Token) or Something you Are (biometrics such as Face, Iris, or Fingerprint). {\it Credential Type } can also include Two-factor authentication, which requires selecting two credential types. 
All credential types must be associated with a {\it level of automation} , i.e. manual (0.0), semi-automated (0.5) or fully automated (1.0).

Particular features include constraints. For example, {\it password } can have three levels of {\it strength}, i.e. low (0.5), medium (0.7), or high (1.0). Credentials can be renewed weekly or monthly, and the selection of a device ({\it Device Types}) depends on the chosen credential type. For example, a reader can only be used when smart card authentication is selected. Finally, some {\it credential types} depend on  {\it cryptography type}  (cryptographic algorithms), i.e., signature, signcryption, group signature, or ring signature.

In the lower section of the figure, we establish connections between contextual factors and the features they render infeasible (-f). For example, low lighting conditions make face and iris recognition ineffective~\cite{wojtowicz2016model}. This can be formalized, as shown below:

\begin{equation}
 LowLighting \implies (Face = 0) \land (Iris = 0)
\end{equation}

Additionally, we associate each feature with its contribution to satisfying different requirements. To represent the impact of each feature on requirement satisfaction, we use qualitative labels: highly positive ($++s$), i.e., determines satisfaction $> 0.8$), positive ($+s$), i.e. determines satisfaction in $(0.6,0.8]$, negative ($-s$), i.e. determines satisfaction in $(0.2,0.4]$, and highly negative ($--s$), i.e. determines satisfaction $\leq 0.2$. If no impact is indicated, it means that a feature determines a satisfaction in $(0.4, 0.6]$ for that requirement. To maintain clarity in Figure~\ref{fig:features}, not all requirement satisfaction impacts are explicitly shown. For instance, {\it Car Plate \&} {\it Driver's License} has a highly positive contribution towards the satisfaction of the {\it Performance} requirement. Conversely, while {\it Certificate}-based authentication has a highly positive impact on the {\it Authenticity} and {\it Confidentiality} requirements, it has a negative impact on the {\it Performance } requirement as it increases the time to perform authentication. Features like{\it Face} and {\it Iris}  recognition positively contribute to {\it Effectiveness} and {\it Efficiency} while also having a positive impact on the {\it Confidentiality} and {\it Integrity} requirements. Since these impacts cannot be precisely quantified, we rely on qualitative assessments based on our experience and prior research (e.g.,~\cite{idrus2013review,kizza2024authentication,reese2019usability}).
  
The satisfaction of a leaf goal is formalized in Eq. 4, where $g$ is a leaf goal, $P(g)$ is its priority, $m_i$ is one of the enacted authentication methods, and $impact(m_i,g)$ indicates the impact of an authentication method towards the satisfaction of goal $g$.

\begin{equation}
S(g) = P(g) \times {\max_{m_i \in M} impact(m_i, g)}
\end{equation}

Note that the {\it Efficiency} requirement can be positively influenced by two alternative factors: (i) the intrinsic efficiency of the selected authentication method (e.g., {\it Iris}, {\it Face}) and (ii) the automation level of the associated credential. Either factor alone can be sufficient to yield high efficiency from the user’s perspective. By taking the maximum, we model a best-case interpretation in which the Efficiency goal is satisfied if at least one of these factors provides high efficiency. In other words, a highly automated credential can compensate for a less efficient authentication method, and conversely, a highly efficient authentication method can offset a lower level of automation.  The investigation of alternative aggregation operators (e.g., conjunctive or additive effects) will be performed in future work.

If two-factor authentication is enabled, we compute the impact as the average contribution of the two enabled authentication methods. Only for the security requirements ({\it Confidentiality}, {\it Integrity}, and {\it Authenticity}), we add a threshold (configured as $0.2$) to the average contribution of the two authentication methods. This represents the extra protection gained from requiring two authentication methods instead of one (See Eq. ~\ref{eq_5}). Since a goal’s satisfaction value cannot exceed $1$, the result is capped at $1$ whenever the computed value is greater than this limit.

\begin{equation}
\label{eq_5}
S(g) = \min({P(g) \times {\sum_{m_i \in M} impact(m_i, g)}/2},1)
\end{equation}

\subsubsection{Secuirty Risk}

\begin{figure}[h!]
	\centerline{
		\includegraphics[width=0.55\columnwidth] {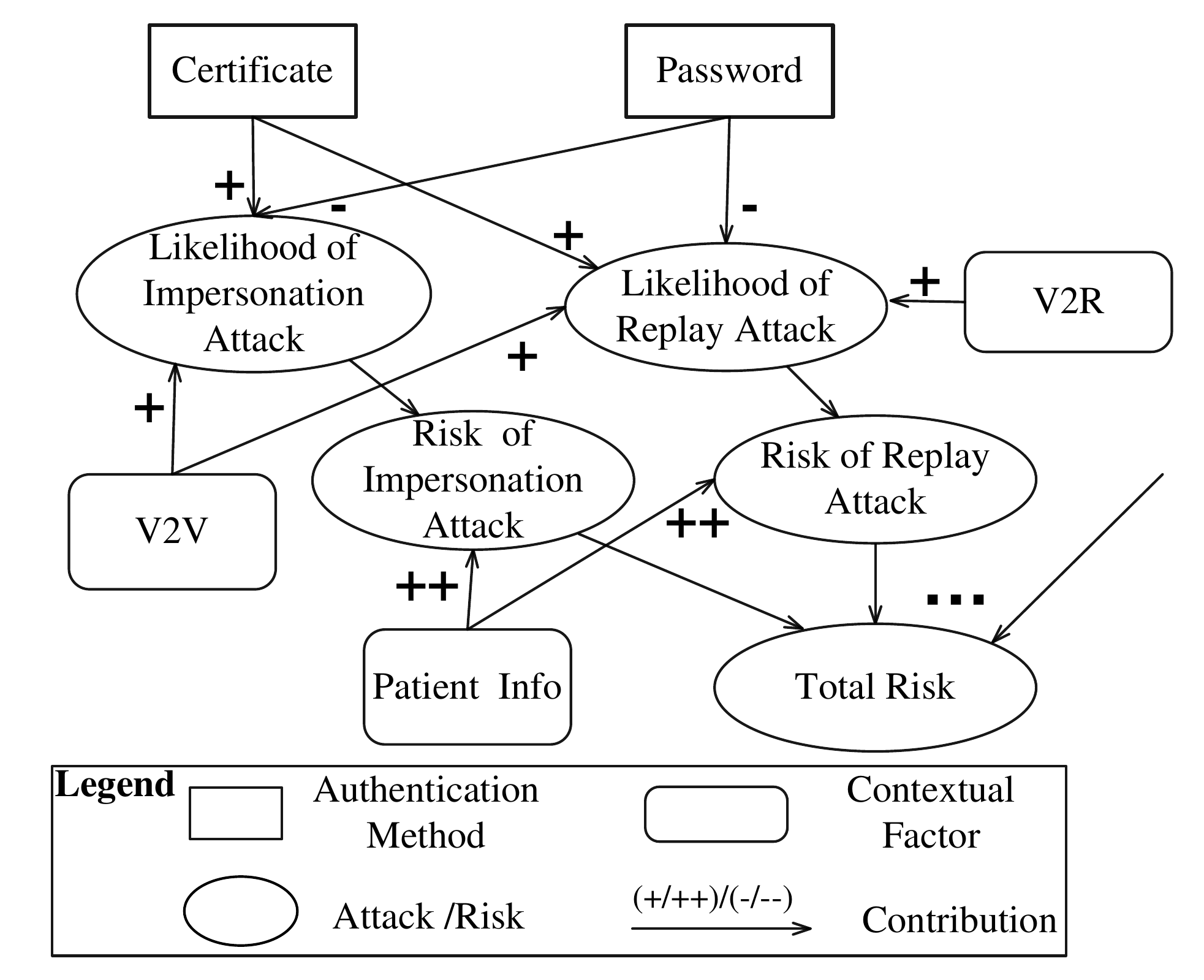}}
	\caption{Security Risks}
	\label{fig:Risks}
\end{figure}

A partial security risk determined by an individual attack depends on the corresponding attack's likelihood and the harm it can cause. An attack likelihood depends on contextual factors and the enabled authentication method. We represent the impact of contextual factors and authentication methods on the likelihood of attacks using qualitative labels. For example, as shown in Figure~\ref{fig:Risks}, the use of a V2R network topology determines a high ($+$) likelihood of an impersonation attack (in $(0.6,0.8]$). Similarly, the use of a V2V network topology determines a high likelihood of impersonation and replay attacks. If more than one contextual factor affects the likelihood of an attack, we adopt a conservative approach and aggregate their impacts, taking the maximum.

Conversely, authentication methods contribute to reducing the likelihood of attacks by a given factor. For example, as shown in Figure~\ref{fig:Risks}, certificate-based authentication highly decreases ($+$) the likelihood of impersonation and replay attacks (by a factor in $(0.6,0.8]$. Password-based authentication mildly decreases ($-$) the likelihood of an impersonation attack  (by a factor in $(0.2,0.4]$). If more than one authentication method is enabled, we aggregate their impacts considering the maximum.

The likelihood of an attack ($a$) is computed as formalized in Eq. 6, where $c_j$ and $m_i$ are the enabled contextual factors and authentication methods, respectively.

\begin{equation}
L(a) = \max \left( 0,\; \max_{c_j \in CF} \text{Ctx}(c_j,a)\;-\; \max_{m_i \in AM} \text{C}(m_i,a) \right)
\end{equation}

The likelihood of an attack is equal to the maximum impact of the enabled contextual factors on the attack ($\text{Ctx}(c_j,a)$) minus the maximum impact that the enabled authentication method has on it ($\text{C}(m_i,a)$). The likelihood of an attack is lower-bounded by $0$ to ensure it is never assigned a negative value.


 Contextual factors can affect the harm caused by an attack. For example, we use the sensitivity of the information shared in the IoV network to estimate the damage that can result from impersonation attacks. For example, as shown in Figure~\ref{fig:Risks}, if patient information is exchanged ({\it Patient Info}), the harm caused by an attack is very high ($++$), i.e. $>~0.8$. If more than one contextual factor affects the harm from an attack, we adopt a conservative approach and aggregate their impacts, taking the maximum.

In summary, the partial risk of a specific attack ($PRisk(a)$) depends on its likelihood multiplied by its harm ($H$), as indicated in Eq. 7. 

\begin{equation}
PRisk(a) = L(a) * \max_{c_j \in CF} \text{H}(c_j,a)
\end{equation}

Finally, we adopt a conservative approach and assign the total risk the maximum value of the partial risk of any attacks, as shown in Eq. 8.

\begin{equation}
TRisk = \max_{a_k \in A} \text{PRisk}(a_k)
\end{equation}

\subsection{Analyze and Plan}
 
 We use a Fuzzy Causal Network (FCN)~\cite{FCN} to analyze the impact of contextual factors on the system goals, requirements, and security risk and identify an effective authentication method.

 \subsubsection{Fuzzy Causal Networks (FCNs)} 
integrate the graph-based formalism of causal models with the expressive power of fuzzy logic. 
FCNs enable causal reasoning under uncertainty by aggregating fuzzy influences through the network. When a subset of nodes (e.g., contextual conditions) is activated or observed, their impact can be propagated through the network to estimate their influence on downstream variables. This is typically done using aggregation functions, yielding a value for each target node. An FCN is suitable for analyzing the consequences of context changes on requirements priorities, feasible authentication methods and security risks and also the impact of specific authentication methods on the satisfaction of system goals and on the likelihood of security risks.

An FCN is defined as a directed graph $G=(V,E)$ where:
\begin{itemize}

\item {$V$ is a set of nodes representing variables or system concepts.}
\item {$E \subseteq V \times V $ is a set of directed edges representing causal influences.}
\item Each edge $e_{i,j} \in E$ is associated with a fuzzy weight $w_{ij} \in [0,1]$  indicating the degree and direction of influence from node $v_{i}$ to node $v_{j}$.
\end{itemize}

\begin{figure}[h]
	\centerline{
		\includegraphics[width=0.6\columnwidth] {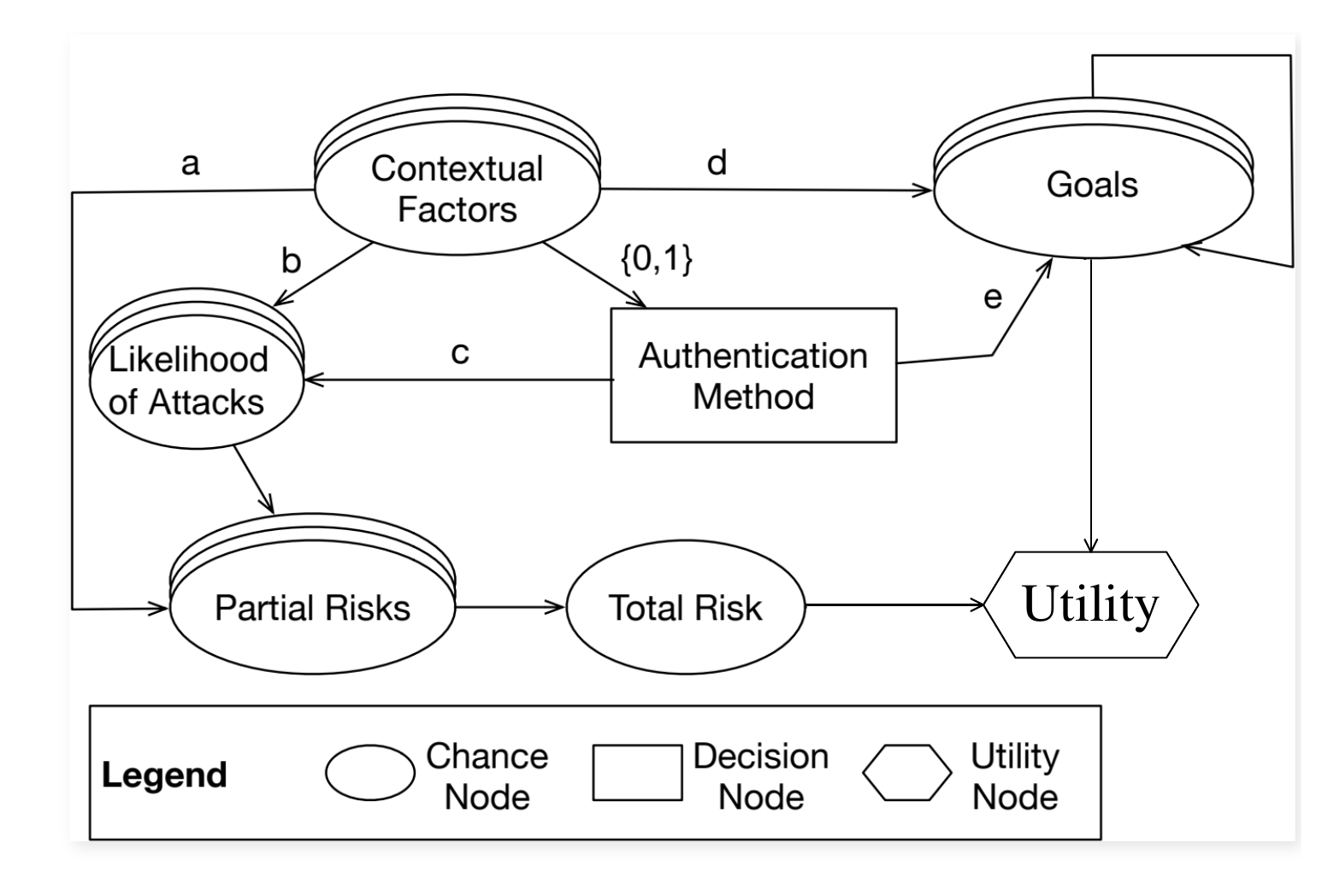}}
	\caption{Abstract Model of the Fuzzy Causal Network}
	\label{fig:FCN}
\end{figure}

Similar to an influence diagram~\cite{InfluenceDiagrams}, an FCN has three types of nodes: chance nodes representing uncertain domain entities significant for causal reasoning (denoted by ovals), decision nodes indicating decisions to be made (denoted by rectangles), and utility nodes corresponding to the fitness value of the network configuration (denoted by a hexagon). All the nodes are represented by variables in the range $[0,1]$. The value of the weights assigned in the FCN  depends on the intensity of the impacts between different entities (e.g., the impact of context on requirements priorities or the likelihood of attacks), as discussed in Section~3.1. We provide an abstract representation of the FCN used in our approach in Figure~\ref{fig:FCN}. The semantics of the weights of the FCN ($a$, $b$, $c$, $d$, $e$) is provided in Table~\ref{tab-3}.



\textbf{Chance nodes.} 
We represent the contextual factors and system goals in Figure~\ref{fig:goals} as chance nodes. The value of contextual factors is monitored at runtime and can influence the priority of goals and requirements as discussed in Section~\ref{sec:goalModel}. As shown in Eq. 2, the value of a non-leaf goal, representing its satisfaction, is determined by the minimum value of its children (see circular influence on the goal node). While, as discussed in Eq. 4, the value of a leaf goal is determined by the influence of enabled contextual factors (represented with weight $d$ in the network) multiplied by the impact of the enabled authentication method(s) ($e$)). Note that if 2-Factor authentication is selected, their average contribution will be considered as shown in Eq. 5.



\begin{table}[htpb]
 	\caption{Semantics of FCN weights. $LF$ - Leaf Goals, $CF$ - Contextual Factors, $At$ - Likelihood of Attacks, $AM$ - Authentication Methods, $PR$ - Partial Risks  }
 	\centering
    \small
 		\begin{tabular}{|p{7cm}|c|}
 			\hline
 			\textbf{Semantics} & \textbf{\textit{ Weight }} \\
 		
 			\hline
 $ pr \in PR,~\max_{c_j \in CF} \text{H}(c_j,pr)$ &  a
 			\\\hline
 	$ a \in At,~~\max_{c_j \in CF} \text{Ctx}(c_j,a)$\ & b\\\hline
 	 $a \in At,~~\max_{m_i \in AM} \text{C}(m_i,a) $ & c\\
    	\hline
        $g \in LF,~c_i \in CF,~~\max_{1 \le i \le N} \operatorname{impact}(c_i, g)$ & d \\\hline
       $\max_{1 \le i \le N} b_i - \max_{1 \le i \le M} c_i $  & e \\
 			\hline

 		\end{tabular}
 		\label{tab-3}

\end{table}


As shown in Eq. 6, contextual factors can positively contribute to the likelihood of attacks ($b$), while authentication methods reduce the likelihood of those attacks ($c$). If more than one contextual factor affects the likelihood of an attack, we aggregate their impact by considering the maximum. The same applies to authentication methods, which reduce the likelihood of an attack. To aggregate these impacts, we sum the positive contribution of the contextual factors with the negative contributions of the selected authentication methods. As shown in Eq. 7, the partial risk of an attack is computed by multiplying its likelihood by the maximum contribution of the enabled contextual factors ($a$). The total risk is computed conservatively as the maximum of the Partial Risks ($h$). Alternative approaches could have been adopted for this purpose, which consider the average of the partial risks.


\textbf{Decision Nodes.} Each authentication method configuration corresponds to a decision
node in the causal network. A decision node represents a specific configuration of the feature model (Figure~\ref{fig:features}), corresponding to the selection of a particular authentication method. As such, a decision node is binary and can take a value of 0 or 1, indicating whether the corresponding authentication configuration is not selected or selected, respectively. This value is then used within the FCN to evaluate the impact of the chosen configuration on security risks and system goals. As shown in Section~\ref{sec:Feature}, authentication methods can reduce the likelihood of attacks, depending on their configuration, and can positively or negatively affect goal satisfaction.  
For example, the strength of a password-based authentication depends on its length, and two-factor authentication is the strongest because it combines the strengths of two selected authentication methods. The stronger a password-based authentication,the higher the impact in reducing the likelihood of an attack.

Contextual factors can also disable specific authentication methods ($\{0,1\}$), thereby influencing the choice of the enabled authentication methods. Thus, if at least one contextual factor negatively affects an authentication method, its value is zero, i.e., it cannot be enabled. For example, when the lighting is low, face and iris recognition cannot be enabled (See Eq. 3).

\textbf{Utility Node.} The utility node expresses the effectiveness
of selected authentication according to the total risk and the
satisfaction of goals. The utility can have a different value whenever a different authentication method is selected. The
utility aggregates the contribution of a chosen authentication method to mitigate the risk and satisfy the goals.  It averages the satisfaction of the root goals ($Security$, $Usability$ and $Performance$) and the contribution of the total risk, by considering its complementary value, as shown in Eq. 9. 

\begin{equation}
Utility = (S(Security) + S(Usability) + S(Performance) + (1-TRisk))/4
\end{equation}

\subsection{Computing an Effective Authentication Method.} 

To compute an effective authentication method, we encode the formalization of the FCN in the language of the Z3 SMT Solver (Z3). According to Z3 notation, all formulae are expressed in the prefix form. For each scenario, we constrain the model with information about the enabled contextual factors. In other words, we assign a value of 1 to the contextual factors present in a particular scenario, while others are assigned a value of 0. For example, in the first scenario (ambulance requiring road traffic information), the contextual factors will be constrained as indicated below, with the V2R Topology enabled and road traffic information exchanged.

\noindent $(assert (= V2Rtopology 1))$\\
$(assert (> RoadTrafficInfo 0))$\\
$(assert (= AmbulanceJunction 0))$\\
$(assert (= PatientInfo 0))$\\
$(assert (= DistanceInfo 0))$\\
$(assert (= NightTime 0))$\\
$(assert (= NodeMovement  0))$\\

We compute the utility by repeatedly solving a satisfiability problem in which the utility is constrained to different candidate values, using binary search over the interval $[0,1]$. The objective of the binary search in this case is to identify the maximum utility. We provide the automated scripts to compute the utility in our repository\footnote{\url{https://github.com/lpasquale/AdaptiveAuthentication}}.

Z3 has been widely evaluated for its scalability in both academic and industrial contexts. Empirical benchmarks from the Satisfiability Modulo Theories Competitions (SMT-COMP)  demonstrate that Z3 can efficiently handle models with up to $100,000$ variables in quantifier-free setting~\cite{SMTComp2023}. While performance diminishes in the presence of quantifiers or non-linear arithmetic, these limitations are typical across most SMT solvers~\cite{de2008z3,Barrett,barrett2011cvc4,barbosa2022cvc5}. Comparative studies indicate that Z3 remains among the top-performing SMT solvers, alongside CVC$5$~\cite{barbosa2022cvc5}, Yices2~\cite{dutertre2014yices}, and MathSAT$5$~\cite{cimatti2013mathsat5}, with Z3 often preferred for its rich API support and integration flexibility~\cite{barbosa2022cvc5}. These properties make it a strong candidate for real-time symbolic reasoning tasks such as those found in adaptive security or authentication systems.
~\cite{Barrett}.
As an alternative, the same analysis could have been expressed as a Constraint Solving Problem (CSP), where the relations among the context, goals, authentication methods,  risks, and utility are represented as constraints, and the objective of the problem is to maximize the utility. However, Z3 is a more mature tool than the available software solutions for CSP problems. In addition, SMT approaches can efficiently solve problems that, at first sight, lack a typical SMT flavour. For example, problems where models are sought such that a given utility function is maximized or a cost function is minimized~\cite{CSP}.

\section{Evaluation}

To evaluate our approach, we assessed its ability to select an effective authentication method based on security risks and the priority of the system's goals. We also evaluated its execution time and the memory usage to assess whether it can be used at runtime. We evaluate the proposed framework using two case studies: Case Study A, which focuses on authentication scenarios in IoV presented in Section~\ref{sec:Scenarios}, and Case Study B, which focuses on authentication scenarios in the healthcare system. These case studies were selected to demonstrate the framework’s applicability across distinct domains and operational conditions.

\subsection{Authentication scenarios in IoV}
\label{sec:IoV_Scenarios}
{ \textbf{Secure Data Exchange (S1)}:} In this scenario, the following contextual factors are enabled, i.e. the vehicle uses a  V2R topology to exchange road traffic information. The impersonation attack also has a high likelihood. The requirements related to the confidentiality of the road traffic information and the authenticity of the parties sharing information (ambulance and RSU) have higher priority than usability and performance requirements. Thus, an effective authentication method shall decrease the risk of an impersonation attack and maximize the satisfaction of the most critical goals (Security).

{\textbf{Ambulance Overtaking Another Vehicle (S2)}:} In this scenario, the following contextual factors are enabled: the authenticating vehicles use a V2V Topology, they exchange distance information, and they are in motion. In this scenario, the impersonation and replay attack are very likely. Also, the performance requirements, Min Auth Time and Min Auth Delay, have higher priority than the Security and Usability requirements. Thus, an effective authentication shall reduce the likelihood of impersonation and reply attack while satisfying the most critical goals (Performance).

Accessing Patient Data at a Junction (S3):  In this scenario, the following contextual factors are enabled, i.e. the ambulance driver is at a junction and accessing patient information using a V2I  cellular network. Because the accessed information is sensitive, maintaining its confidentiality is highly important. Usability requirements are also important since the authentication method should not distract the driver. Thus, an effective authentication shall maximize the satisfaction of the most critical goals (Security and Usability).

\subsubsection{Comparison with the Baseline}

\begin{figure*}[h]
 	\centering
 	\begin{subfigure}[b]{0.45\linewidth}
 		\includegraphics[width=\linewidth]{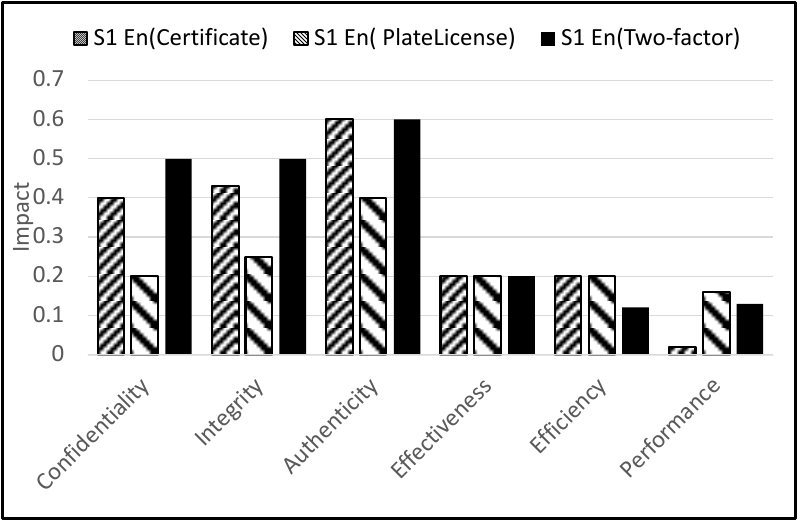}
 		\caption{Secure Data Exchange (S1)}
 		\label{fig-8a}
 	\end{subfigure}
 	\begin{subfigure}[b]{0.45\linewidth}
 		\includegraphics[width=\linewidth]{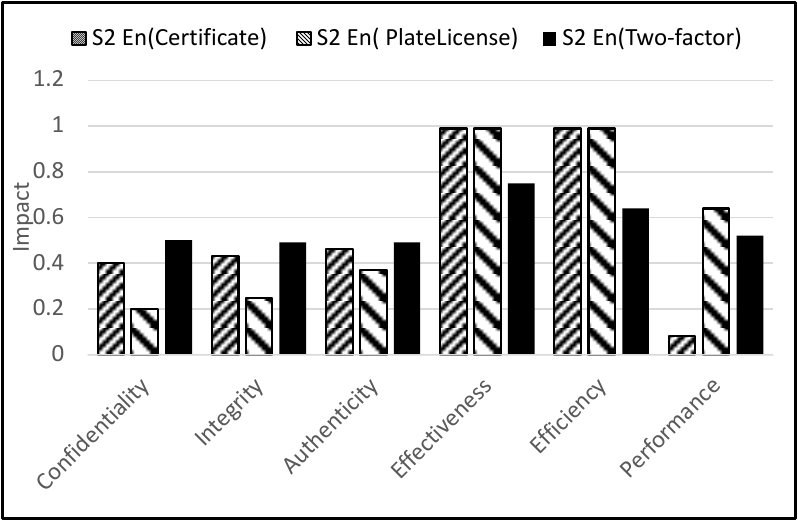}
 		\caption{Ambulance Overtaking Another Vehicle (S2)}
 		\label{fig-8b}
 	\end{subfigure}
 \begin{subfigure}[b]{0.45\linewidth}
 	\includegraphics[width=\linewidth]{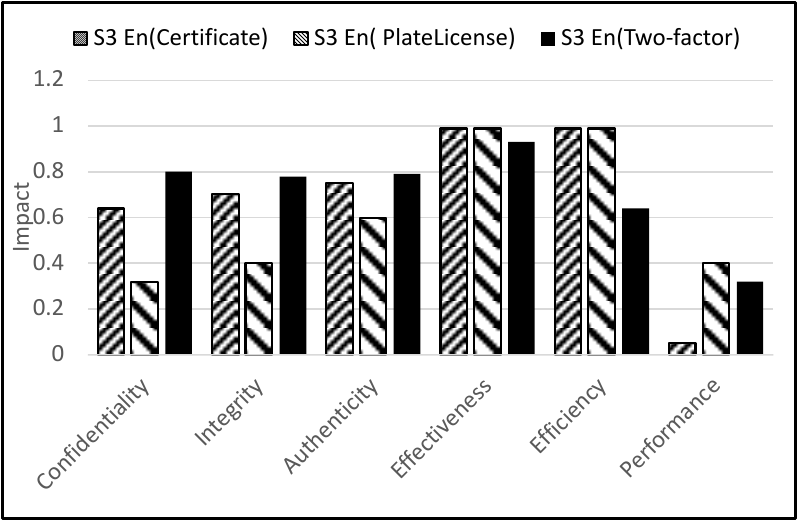}
 	\caption{Accessing Patient Data at a Junction (S3)}
 	\label{fig-8c}
 \end{subfigure}

 	\caption{Impact of the authentication method on the satisfaction of the requirements}
 	\label{fig-8}
 \end{figure*}

We evaluate the proposed framework against a baseline approach that enforces the same authentication method (e.g., certificate-based authentication) across all IoV scenarios, regardless of the operating or contextual environment. We examine how enforcing the same authentication method affects the satisfaction of system requirements across different scenarios. We then compare this baseline with our approach, which dynamically selects authentication methods to satisfy security, usability, and performance requirements under varying IoV conditions.
Figure \ref{fig-8} illustrates the impact of the authentication methods on the satisfaction of security (confidentiality, integrity, and authenticity), usability (effectiveness and efficiency), and performance goals across the evaluated IoV scenarios. The results show that enforcing a single authentication method across all scenarios fails to consistently satisfy the requirements, as each scenario exhibits distinct contextual characteristics and differing requirement priorities. For instance, certificate-based authentication is well-suited to Scenario S1, where confidentiality and authenticity are prioritized, and it also achieves high satisfaction for the remaining requirements. However, this method is not effective in other scenarios, where differing contextual factors and requirement priorities necessitate alternative authentication methods. In S1, the security requirements have high priority. Certificate-based authentication and Two-Factor authentication have a better impact on the satisfaction of the Security goals in S1 compared to the Car Plate \& Driver License (see figure~\ref{fig-8a}). In S2, where performance requirements have higher priority, this authentication method (Certificate) does not ensure reasonable satisfaction of the Performance and Usability goals, and the Care Plate and Driver License (PlateLicense) have a greater impact on their satisfaction (see Figure~\ref{fig-8b}). In S3, the Usability and Security goals have high priority, and Two-Factor authentication has a better overall impact on the satisfaction of these goals (see figure~\ref{fig-8c}). 
Thus, our approach selects certificate-based authentication for S1, car plate and driver license authentication for S2, and two-factor authentication (car plate and driver license verification with fingerprint recognition) for S3. As shown in Figure \ref{fig-8}, these authentication methods maximize the overall satisfaction of security, usability, and performance requirements while reducing security risks. These results demonstrate that incorporating contextual awareness into authentication selection is essential for achieving balanced and effective requirement satisfaction in the IoV environments.

\begin{figure*}[h]
 	\centering
 	\begin{subfigure}[b]{0.45\linewidth}
 		\includegraphics[width=\linewidth]{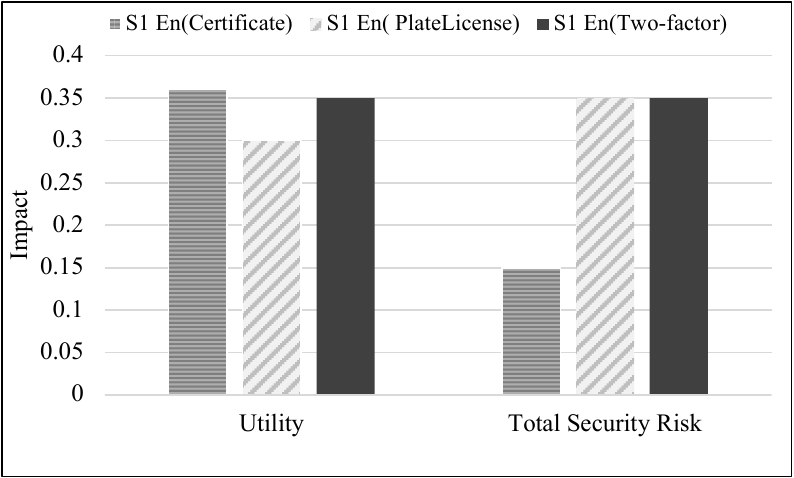}
 		\caption{Secure Data Exchange (S1)}
 		\label{fig-9a}
 	\end{subfigure}
 	\begin{subfigure}[b]{0.45\linewidth}
 		\includegraphics[width=\linewidth]{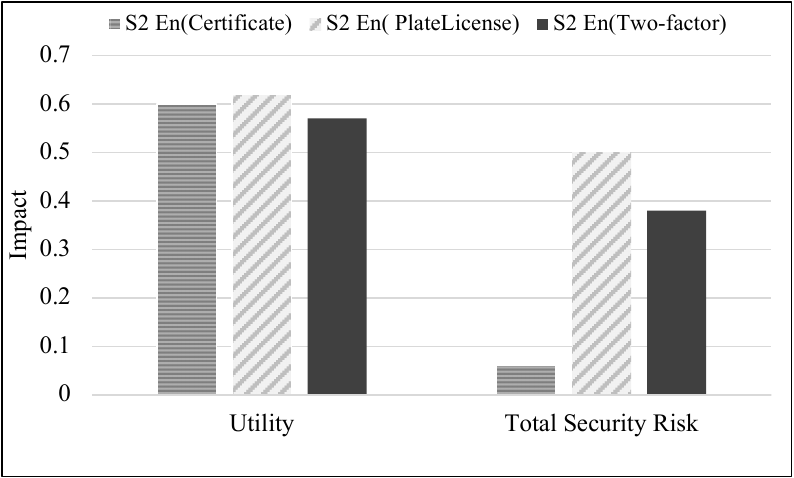}
 		\caption{Ambulance Overtaking Another Vehicle (S2)}
 		\label{fig-9b}
 	\end{subfigure}
 \begin{subfigure}[b]{0.45\linewidth}
 	\includegraphics[width=\linewidth]{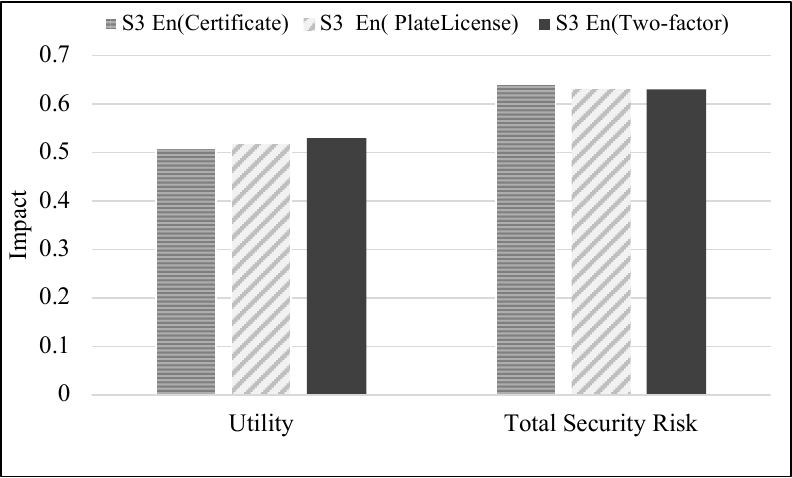}
 	\caption{Accessing Patient Data at a Junction (S3)}
 	\label{fig-9c}
 \end{subfigure}

 	\caption{Impact of the authentication method on mitigating the total risk and maximizing the Utility}
 	\label{fig-9}
 \end{figure*}

We are also interested in understanding how enforcing the same authentication method (e.g., Certificate) mitigates the security risk and maximizes the utility, indicating the tradeoffs between the satisfaction of system goals and the reduction of the total risk across all the IoV scenarios. We then compare our approach with this baseline. The results shown in Figure~\ref{fig-9} demonstrate that enforcing the same authentication method across all scenarios fails to appropriately mitigating the security risk and maximizing the utility, as each scenario exhibits distinct contextual characteristics and requirement priorities. For example, enforcing Certificate-based authentication could mitigate security risks and maximize utility in S1, but will not have a similar impact in the other scenarios (see Figures~\ref{fig-9b} and ~\ref{fig-9c}).  Our approach enforced secure authentication methods (Certificate-based and Two-Factor authentication) when the total risk was higher in S1 and S3(see Figures~\ref{fig-9a} and ~\ref{fig-9c}). In S2, where the total risk was lower, our approach selected the Car Plate and Driver License authentication (Figure~\ref {fig-9b}), which ensured higher satisfaction of the Performance goal (see Figure~\ref{fig-9b}). We learned that although an authentication method mitigates certain security risks and maximizes the utility for a particular scenario (e.g., S1), it does not guarantee that it is effective in another scenario characterized by different contextual factors and security risks. 
These results demonstrate that our approach selects an authentication method that mitigates security risks and maximizes the utility compared to the baseline, in which we choose the same authentication method regardless of contextual factors and requirement priorities.


Changing contextual factors at runtime can affect security risks, priority requirements, and the feasibility of authentication methods. Table~\ref{tab-2} depicts the configuration of contextual factors and their impacts on the priority of the system goals for S1, S2, and S3 measured as high (H), medium (M) and low (L). In S1, using V2R Topology and exchanging road traffic information affects the priority of security, usability and performance goals, which are high, low and medium, respectively. Table~\ref{tab-4} also shows the authentication methods selected by our approach across different scenarios. Our approach chooses certificate-based authentication because it better satisfies the authenticity requirement related to the impersonation attack and also provides reasonable satisfaction of the usability goal, which has medium priority. In S2, using the V2V topology, exchanging distance information between two vehicles, and the fact that cars authenticate while moving, give high priority to the performance goals. Security requirements have a medium priority because the exchanged information is not sensitive, while the usability goal has a low priority (L). Thus, our approach chooses Car Plate and Driver License authentication for S2 because it achieves excellent performance while satisfying the security goals. In S3, the ambulance driver is at a junction, accessing patient information via a Vehicle-to-Infrastructure cellular network. This makes all security, usability, and performance goals have a high priority (H). Thus, our approach chooses a two-factor authentication (Car Plate and Driver License + Fingerprint) because it provides a good balance among security, performance, and usability goals.
\begin{table}[h]
\centering
\Huge 
\renewcommand{\arraystretch}{1.3} 
\caption{Our approach result for Case Study A \& Configuration of Requirements Priorities (HIGH: H, MEDIUM: M, LOW: L)}
\label{tab-4}
\resizebox{\textwidth}{!}{
\begin{tabular}{|>{\centering\arraybackslash}p{0.8cm}|
                >{\centering\arraybackslash}p{5.7cm}|
                >{\centering\arraybackslash}p{2.2cm}|
                >{\centering\arraybackslash}p{2.5cm}|
                >{\centering\arraybackslash}p{3.4cm}|
                >{\centering\arraybackslash}p{3.1cm}|
                >{\centering\arraybackslash}p{3.9cm}|
                >{\centering\arraybackslash}p{3.7cm}|
                >{\centering\arraybackslash}p{3.2cm}|}
\hline
\multirow{2}{*}{\textbf{S}} & 
\multirow{2}{*}{\textbf{Contextual Factors}} & 
\multicolumn{3}{c|}{\textbf{Requirements Priorities}} & 
\multicolumn{4}{c|}{\textbf{Enabled Authentication Method}} \\ 
\cline{3-9} 
& & \textbf{Security} & \textbf{Usability} & \textbf{Performance} & 
\textbf{Something You Know} & \textbf{Something You Have} & 
\textbf{Something You Are} & \textbf{Two-Factor} \\ 
\hline
\textbf{S1} & \makecell{V2RTopology = 1; \\ RoadTraffic = 1} & H & L & M & \textemdash & En(Certificate) & \textemdash & \textemdash \\ 
\hline
\textbf{S2} & \makecell{V2VTopology = 1; \\ DistanceInfo = 1; 
\\ NodeMovement = 1} & M & L & H & \textemdash & En(PlateLicense) & \textemdash & \textemdash \\ 
\hline
\textbf{S3} & \makecell{V2ITopology = 1; \\ PatientInfo=1; 
\\ AmbulanceJunction = 1} & H & H & H & \textemdash & En(PlateLicense) & En(Fingerprint) & En(1) \\ 
\hline

\end{tabular}%
}
\label{tab-2}
\end{table}

\subsubsection{Overhead}
To assess the overhead of our approach in Internet of Vehicles (IoV) scenarios, we computed the authentication method for each scenarios 100 times, and measured the minimum, maximum, and average were  time and memory overhead. To do the experiments we used a MacBook Pro  Apple M2 Pro chip, with 8 cores and 8 GB of RAM. Table~\ref{tab:IoV-runtime} shows the time required to identify an authentication method and the associated memory usage.  Notably, the models demonstrated the ability to generate a ranked list of configurations in under 1.52 seconds for each scenario while utilizing less than 19 MB of memory. These results indicate that this type of analysis and planning activities necessary to identify an effective authentication method can be performed at runtime.

  \begin{table}[htbp]
\centering
\caption{Runtime and memory usage for the IoV scenarios.}
\begin{tabular}{|c|c|c|c|c|c|c|c| }
\hline
Scenario & Runs & Time$_{\min}$ & Time$_{\text{avg}}$ & Time$_{\max}$ & Memory$_{\min}$ & Memory$_{\text{avg}}$ & Memory$_{\max}$ \\
& & (ms) & (ms) & (ms) & (MB) & (MB) & (MB) \\
\hline
1 & 100 & 904.46 & 962.42 & 1473.82 & 9.97 & 13.68 & 18.98  \\
\hline
2 & 100 & 1228.51 & 1277.45 & 1519.90 & 9.78 & 11.59 & 12.78 \\
\hline
3 & 100 & 461.95  & 484.58  & 725.17  & 9.98 & 12.09 & 12.64 \\
\hline
\end{tabular}
\label{tab:IoV-runtime}
\end{table}


To evaluate the complexity of the SMT problems solved for the IoV scenarios, we analyze the FCN model size and the corresponding solver effort in Z3 for each scenario (see Table~\ref{tab:iov-smt-metrics}).
For the model size, we report: (i) the number of declared variables (i.e., declare-fun and declare-const statements), (ii) the number of asserted constraints (assert statements), (iii) the number of arithmetic expressions (arith-terms), and (iv) the number of conditional expressions (ite-terms).
For the solver effort, we consider standard Z3 solving metrics, including the number of branching decisions, propagated literals, and solver restarts. While declarations and assertions characterize the structural size and constraint density of the model, decisions, propagations, and restarts reflect the computational effort required by the solver to identify a satisfying configuration.
Taken together, these metrics indicate both the runtime feasibility and complexity of our approach, demonstrating that the model can be executed online to support runtime adaptation decisions.

  \begin{table}[htbp]
\centering
\caption{SMT formula size and solver effort for IoV scenarios.}
\begin{tabular}{|c|c|c|c|c|c|c|c| }
\hline
Scenario 
& \#variables 
& \#asserts 
& \#arith-terms 
& \#ite-terms 
& \#decisions 
& \#propagations 
& \#restarts \\
\hline
S1 & 86 & 172 & 151 & 115 & 1244 & 8045 & 1\\
\hline
S2 & 86 & 171 & 151 & 115 & 9930 & 20498 & 3\\
\hline
S3 & 86 & 171 & 151 & 115 & 2016 & 5015 & 1\\

\hline
\end{tabular}
\label{tab:iov-smt-metrics}
\end{table}

 \subsection{Authentication Scenarios in Healthcare System } \label{healthcare_scenarios}
We apply our framework to the healthcare domain to assess its generalizability. Healthcare information systems manage highly sensitive, safety-critical patient data across interconnected components, including electronic health records, prescription platforms, and administrative systems. These systems are accessed by multiple users with different roles, including clinicians and administrative staff, operating under dynamic conditions such as shared workstations, remote access, shift changes, and emergency situations \cite{7238590} (See Fig.~\ref{fig:healthcare_scenario}). Such contextual variations introduce authentication-related risks \cite{khan_systematic_2023} (e.g., impersonation and session-hijacking attacks), requiring authentication mechanisms that mitigate security threats while maintaining strong usability and performance guarantees. The combination of sensitive data, complex workflows, and rapidly changing operational contexts makes the healthcare domain a compelling case study for evaluating adaptive, context-aware authentication approaches. Similarly to the IoV case study, we evaluate our adaptive authentication framework against a baseline that enforces a single authentication method across varying representative healthcare scenarios, described below:


 \begin{figure*}[h]
 	\centering
 	\begin{subfigure}[b]{0.33\linewidth}
 		\includegraphics[width=\linewidth]{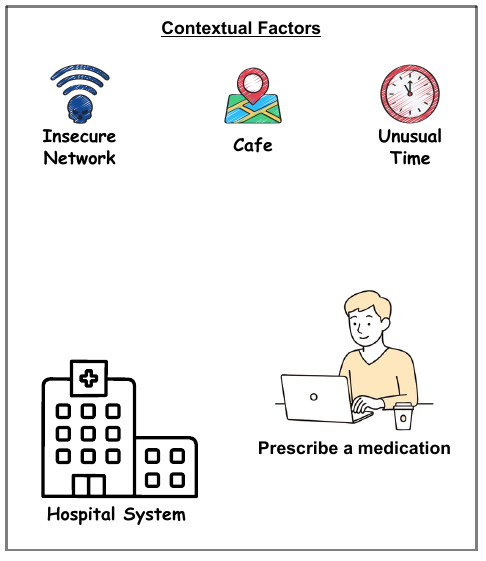}
 		\caption{Outside the Hospital (S4)}
 		\label{fig-10a}
 	\end{subfigure}
 	\begin{subfigure}[b]{0.33\linewidth}
 		\includegraphics[width=\linewidth]{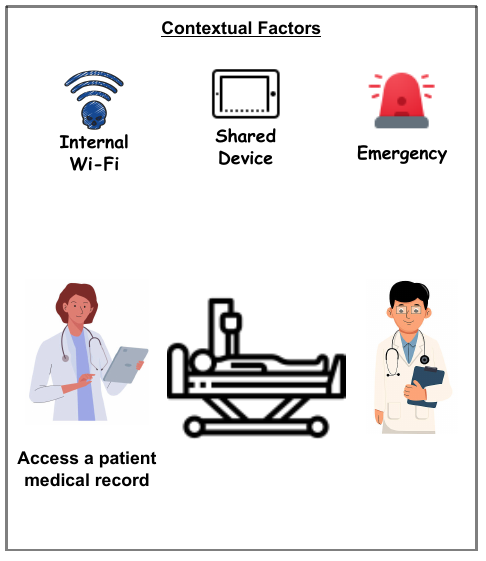}
 		\caption{Doctor in Emergency Department (S5)}
 		\label{fig-10b}
 	\end{subfigure}
 \begin{subfigure}[b]{0.33\linewidth}
 	\includegraphics[width=\linewidth]{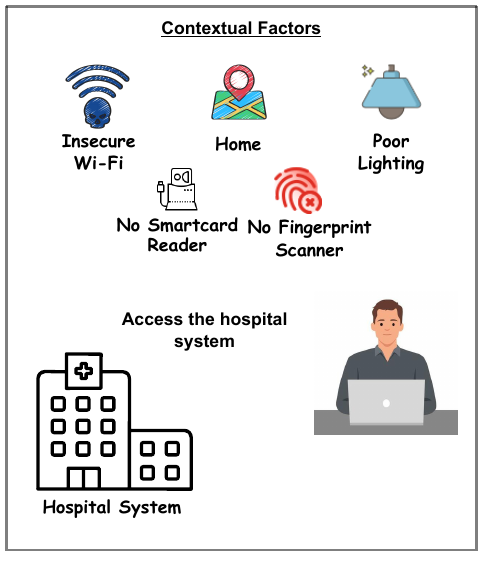}
 	\caption{Late Night Access (S6)}
 	\label{fig-10c}
 \end{subfigure}
 	\caption{Adaptive Authentication Scenarios in Healthcare.}
 	\label{fig:healthcare_scenario}
 \end{figure*}

\textbf{Outside the Hospital (S4)}: In this scenario, a family physician attempts to prescribe medication outside the hospital from an unknown location, over an insecure network, and at an unusual time. In this scenario, impersonation and session‑hijacking attacks are highly likely. Also, confidentiality and integrity of the prescribing action have higher priority than usability and performance goals. Thus, an effective authentication method shall reduce the likelihood of impersonation and session‑hijacking attacks and maximize the satisfaction of confidentiality and integrity requirements.

\textbf{Doctor in the Emergency Department (S5)}: In this scenario, a physician in the emergency department needs to authenticate quickly to access a patient’s medical records during a critical situation using a hospital tablet connected to the internal Wi-Fi network. Because the environment is crowded and devices are shared among staff, impersonation and replay attacks are likely. Due to the emergency, the performance requirement has a higher priority than the security requirements. Thus, an effective authentication method shall reduce the likelihood of impersonation and replay attacks while satisfying the most critical goals (Performance).

\textbf{Late Night Access (S6)}: In this scenario, the physician accesses the hospital system from home at night using a personal, unmanaged laptop on an unsecured Wi‑Fi network. The device lacks a fingerprint scanner, and the smart‑card reader is unavailable. Because access occurs from an untrusted device and over an insecure connection, the risk of session hijacking is high. Therefore, the security requirement has the highest priority in this context. An effective authentication method shall reduce the likelihood of session‑hijacking (and related replay) attacks to satisfy security goals, which are the most critical in this scenario. We provide the automated scripts to replicate our approach of healthcare scenarios in our repository\footnote{\url{https://github.com/Kabashi01/Adaptive-Authentication}}.


\subsubsection{Comparison with the Baseline}

As in the IoV case study (Section ~\ref{sec:IoV_Scenarios}), we compare our approach against a baseline that enforces the same authentication method across all healthcare scenarios, independent of contextual factors.

\begin{figure*}[h]
 	\centering
 	\begin{subfigure}[b]{0.45\linewidth}
 		\includegraphics[width=\linewidth]{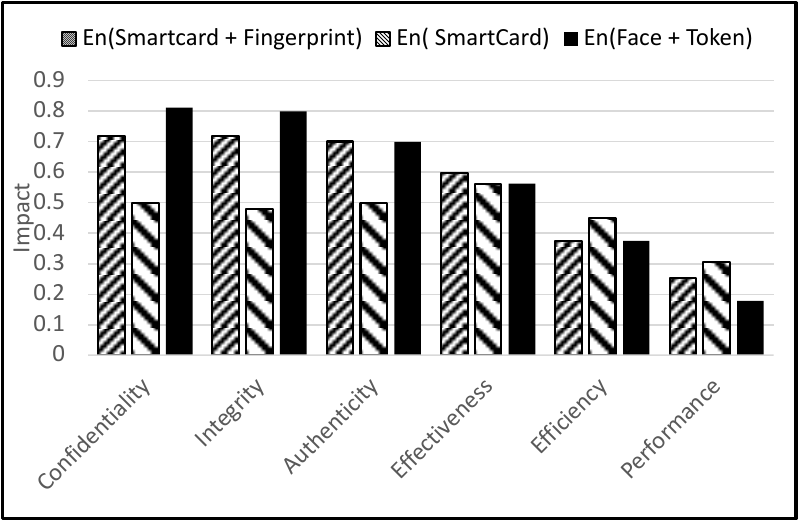}
 		\caption{Outside the Hospital (S4)}
 		\label{fig-11a}
 	\end{subfigure}
 	\begin{subfigure}[b]{0.45\linewidth}
 		\includegraphics[width=\linewidth]{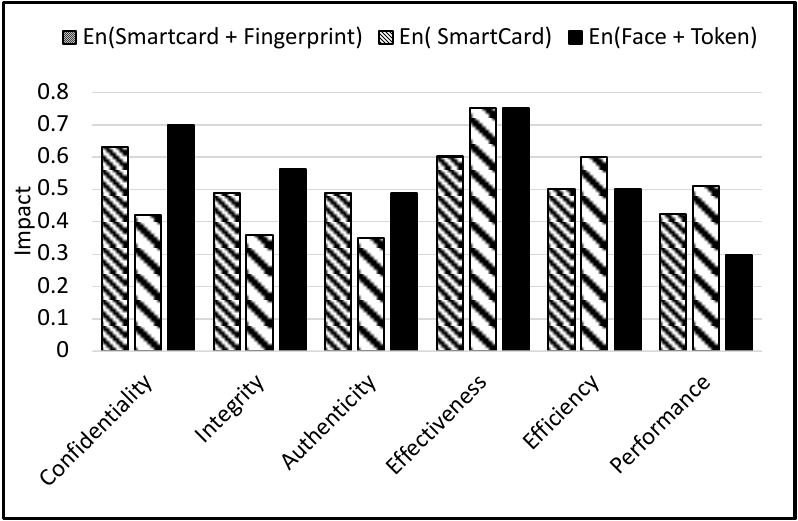}
 		\caption{Doctor in the Emergency Department (S5)}
 		\label{fig-11b}
 	\end{subfigure}
 \begin{subfigure}[b]{0.45\linewidth}
 	\includegraphics[width=\linewidth]{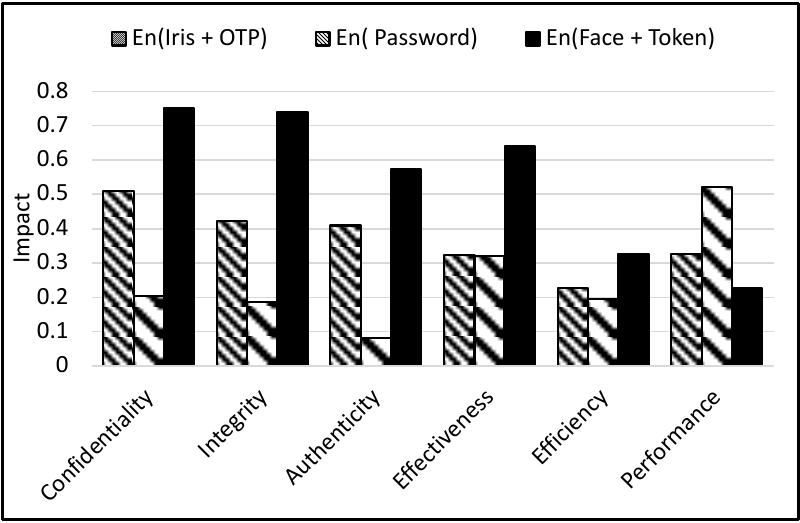}
 	\caption{Late Night Access (S6)}
 	\label{fig-11c}
 \end{subfigure}

 	\caption{Impact of the authentication method on the satisfaction of the requirements (Healthcare Scenarios)}
 	\label{fig-11}
 \end{figure*}

Figure \ref{fig-11} illustrates the impact of the authentication methods on the satisfaction of security (confidentiality, integrity, and authenticity), usability (effectiveness and efficiency), and performance goals across the evaluated healthcare scenarios. The results show that enforcing the same authentication method (e.g., smartcard) across all scenarios fails to consistently satisfy the goals, as each scenario exhibits distinct contextual characteristics and requirement priorities. For instance,  smartcard-based authentication is well-suited to Scenario S5, where performance is prioritized, and it also achieves high satisfaction for the remaining requirements. However, this method is not the effective choice in other scenarios, where differing contextual constraints and requirement priorities necessitate alternative authentication methods. 
In S4, where security goals have higher priority, smartcard-based authentication does not ensure a good satisfaction of these goals, and the Two-factor authentication (SmartCard + Fingerprint) has a better impact on the satisfaction of these goals (see Figure~\ref{fig-11b}). In S6, the Usability and Security goals have high priority, and Two-Factor authentication (Token + Face) has a better overall impact on the satisfaction of these goals (see figure~\ref{fig-11c}) than smartcard-based authentication. 
As shown in Figure \ref{fig-11}, our approach selection enables our framework to better maximize the overall satisfaction of security, usability and performance requirements. These results demonstrate that incorporating contextual awareness into authentication selection is essential for achieving balanced and effective requirement satisfaction in the presented healthcare scenarios.

\begin{figure*}[h]
 	\centering
 	\begin{subfigure}[b]{0.45\linewidth}
 		\includegraphics[width=\linewidth]{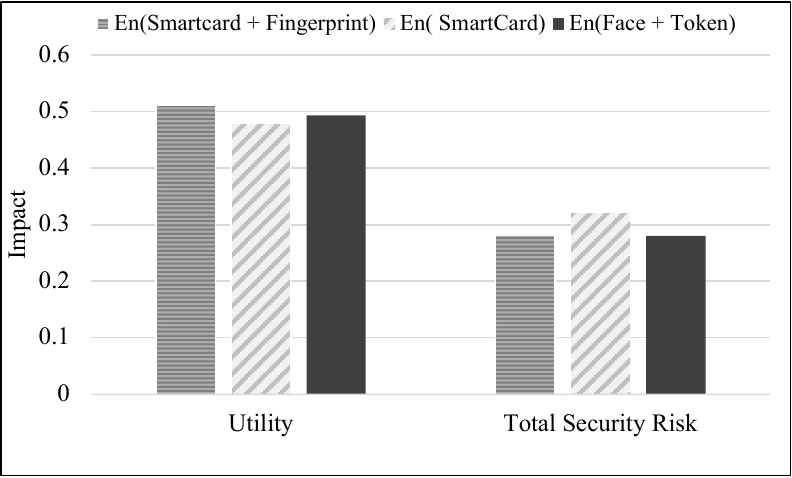}
 		\caption{Outside the Hospital (S4)}
 		\label{fig-12a}
 	\end{subfigure}
 	\begin{subfigure}[b]{0.45\linewidth}
 		\includegraphics[width=\linewidth]{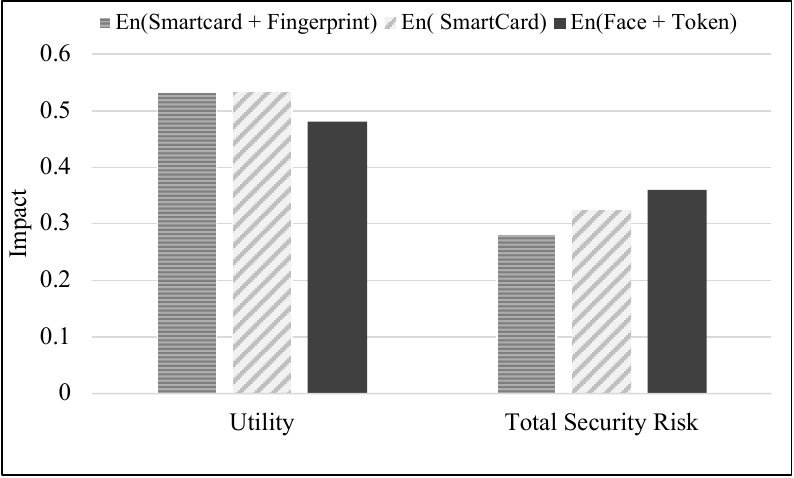}
 		\caption{Doctor in the Emergency Department (S5)}
 		\label{fig-12b}
 	\end{subfigure}
 \begin{subfigure}[b]{0.45\linewidth}
 	\includegraphics[width=\linewidth]{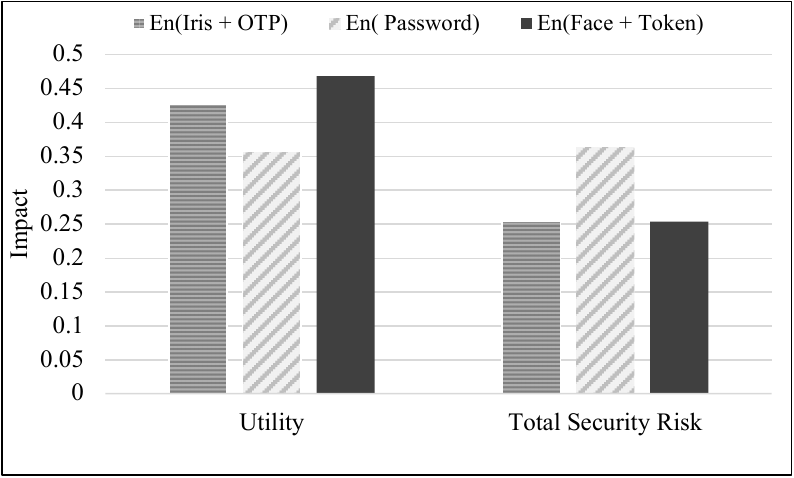}
 	\caption{Late Night Access (S6)}
 	\label{fig-12c}
 \end{subfigure}

 	\caption{Impact of the authentication method on mitigating the total risk and maximizing the utility (Healthcare Scenarios).}
 	\label{fig-12}
 \end{figure*}

We further investigate the extent to which enforcing the same authentication method mitigates security risks and maximizes FCN utility—by balancing system goals and risk reduction—across diverse healthcare scenarios. We then compare this baseline with our proposed approach. Figure~\ref{fig-12} shows the impact of choosing the authentication methods on mitigating the total risk and maximizing the utility. The results show that enforcing the same authentication method across all scenarios fails to mitigate the security risk and maximizing the utility, as each scenario exhibits distinct contextual characteristics and requirement priorities. For example, enforcing smartcard-based authentication mitigates security risks and maximizes utility in S5, but does not achieve comparable benefits in the other scenarios (see Figures~\ref{fig-12a} and~\ref{fig-12c}). When the overall risk is higher, as in S4 and S6, our approach enforces more secure authentication mechanisms, specifically two-factor authentication (see Figures~\ref{fig-12a} and~\ref{fig-12c}). In contrast, in S5—where the total risk is lower—it selects an authentication method that better supports the Performance goal (see Figure~\ref{fig-12b}).

These results show that an authentication method that effectively mitigates security risks and maximizes utility in one scenario may not be equally effective in another due to differences in contextual factors and risk profiles. Similarly, the two-factor authentication configuration (SmartCard + Fingerprint) chosen in S4 is less effective when applied to S5 and S6. In S5, the approach selects smartcard-based authentication (Figure~\ref{fig-12b}), which provides the highest utility by balancing security and performance requirements. In S6, it selects two-factor authentication (Token + Face) (Figure~\ref{fig-12c}), as it most effectively reduces security risks while maximizing overall utility.

Overall, these findings demonstrate that our approach dynamically selects authentication methods that balance security risk mitigation and requirement satisfaction more effectively than the baseline, which applies a single authentication method irrespective of contextual factors and requirement priorities.

Changes in contextual factors at runtime can significantly influence security risks, requirement priorities, and the feasibility of available authentication methods. Table~\ref{tab-4} summarizes the contextual configurations and their impact on system goal priorities for scenarios S4, S5, and S6, classified as high (H), medium (M), or low (L).

In S4, the use of an unknown location, an unusual access time, and an insecure network increase the security goal's priority to high, while usability and performance are assigned low priority. As shown in Table~\ref{tab-4}, our approach selects two-factor authentication (SmartCard + Fingerprint) for this scenario, as it best satisfies the high-priority security requirement while still providing acceptable usability and performance, both of which have lower priority.

In S5, the presence of a shared device and an emergency situation elevates the performance goal's priority to high. Security is assigned medium priority due to the use of an internal hospital Wi-Fi network and managed devices, while usability also has medium priority. Consequently, our approach selects smartcard-based authentication, which offers excellent performance while maintaining sufficient security and usability.

In S6, the physician accesses the hospital system from home at night using an unmanaged laptop on an unsecured Wi-Fi network. The device lacks a fingerprint scanner, and a smartcard reader is unavailable. Under these conditions, security goals have the highest priority, usability is assigned medium priority, and performance has low priority. Accordingly, our approach selects two-factor authentication (Token + Face), as it provides the best balance among security, usability, and performance goals in this context.

\begin{table}[h]
\centering
\Huge 
\renewcommand{\arraystretch}{1.3} 
\caption{Our approach result for Case Study B \& Configuration of Requirements Priorities (HIGH: H, MEDIUM: M, LOW: L)}
\label{tab-healthcare_context}
\resizebox{\textwidth}{!}{
\begin{tabular}{|>{\centering\arraybackslash}p{0.8cm}|
                >{\centering\arraybackslash}p{5.7cm}|
                >{\centering\arraybackslash}p{2.2cm}|
                >{\centering\arraybackslash}p{2.5cm}|
                >{\centering\arraybackslash}p{3.4cm}|
                >{\centering\arraybackslash}p{3.1cm}|
                >{\centering\arraybackslash}p{3.9cm}|
                >{\centering\arraybackslash}p{3.7cm}|
                >{\centering\arraybackslash}p{3.2cm}|}
\hline
\multirow{2}{*}{\textbf{S}} & 
\multirow{2}{*}{\textbf{Contextual Factors}} & 
\multicolumn{3}{c|}{\textbf{Requirements Priorities}} & 
\multicolumn{4}{c|}{\textbf{Authentication Methods}} \\ 
\cline{3-9} 
& & \textbf{Security} & \textbf{Usability} & \textbf{Performance} & 
\textbf{Something You Know} & \textbf{Something You Have} & 
\textbf{Something You Are} & \textbf{Two-Factor} \\ 
\hline
\textbf{S4} & \makecell{UnknownLocation = 1; \\ UnusualTime = 1; \\ InsecNetwork = 1} & H & L & L & \textemdash & En(Smartcard) & En(Fingerprint) & En(1) \\ 
\hline
\textbf{S5} & \makecell{SharedDevice = 1; \\ Emergency = 1} & M & M & H & \textemdash & En(Smartcard) & \textemdash & \textemdash \\ 
\hline
\textbf{S6} & \makecell{Location = 1; \\ UnknownDevice=1; 
\\ UnsecuredWiFi = 1; \\ UnusualTime = 1; \\ Reader = 0; Scanner = 0} & H & M & L & \textemdash & En(Token) & En(Face) & En(1) \\ 
\hline

\end{tabular}%
}
\label{tab-4}
\end{table}


\subsubsection{Overhead}
To assess the overhead of our approach in the healthcare scenarios, we reused the same experimental setup as for the IoV scenarios in section 6.1.  Table~\ref{tab:healthcare-runtime} summarizes the execution time and memory overhead of the healthcare scenarios (S4–S6). For S4, the solver requires 2814.98-4513.09ms per run, with an average of 3007.5ms. For S5, the execution time ranges from 2279.45 ms to 2789.85 ms, with an average of 2368.51 ms. In contrast, S6 is significantly faster, with a runtime between 340.35 ms and 422.45 ms, and an average of 358.64 ms. Memory usage remains relatively stable across all scenarios, varying between approximately 7.83 MB and 15.45 MB (with averages of 10.43 MB for S4, 9.80 MB for S5, and 11.27 MB for S6), indicating that the observed performance differences are mainly driven by the size of the search space rather than raw memory consumption. The lower execution time of S6 compared to S4 and S5 is mainly due to contextual factors that make both the scanner and the smartcard reader unavailable, thereby disabling two authentication methods (Smartcard and Fingerprint) and significantly reducing the search space for an optimal authentication configuration. This suggests that the number of enabled authentication methods in the adaptive authentication model significantly affects the time required to find the optimal authentication configuration.  



 \begin{table}[th]
\centering
\caption{Runtime and memory usage for healthcare scenarios.}
\begin{tabular}{|c|c|c|c|c|c|c|c| }
\hline
Scenario & Runs & Time$_{\min}$ & Time$_{\text{avg}}$ & Time$_{\max}$ & Memory$_{\min}$ & Memory$_{\text{avg}}$ & Memory$_{\max}$ \\
& & (ms) & (ms) & (ms) & (MB) & (MB) & (MB) \\
\hline
4 & 100 & 2814.98 & 3007.50 & 4513.09 & 9.17 & 10.43 & 15.45  \\
\hline
5 & 100 & 2279.45 & 2368.51 & 2789.85 & 7.83 & 9.80 & 11.22 \\
\hline
6 & 100 & 340.35  & 358.64  & 422.45  & 10.58 & 11.27 & 11.86 \\
\hline
\end{tabular}
\label{tab:healthcare-runtime}
\end{table}


To evaluate the complexity of the SMT problems solved for the healthcare scenarios, we analyze the FCN model size and the corresponding solver effort in Z3 for each scenario (see Table~\ref{tab:healcare-smt-metrics}).
For the model size, we report: (i) the number of declared variables (i.e., declare-fun and declare-const statements), (ii) the number of asserted constraints (assert statements), (iii) the number of arithmetic expressions (arith-terms), and (iv) the number of conditional expressions (ite-terms).
For the solver effort, we consider standard Z3 solving metrics, including the number of branching decisions, propagated literals, and solver restarts. While declarations and assertions characterize the structural size and constraint density of the model, decisions, propagations, and restarts reflect the computational effort required by the solver to identify a satisfying configuration.
Taken together, these metrics indicate both the runtime feasibility and complexity of our approach, demonstrating that the model can be executed online to support runtime adaptation decisions.

  \begin{table}[htbp]
\centering
\caption{SMT formula size and solver effort for healthcare scenarios.}
\begin{tabular}{|c|c|c|c|c|c|c|c| }
\hline
Scenario 
& \#variables 
& \#asserts 
& \#arith-terms 
& \#ite-terms 
& \#decisions 
& \#propagations 
& \#restarts \\
\hline
S4 & 90 & 178 & 177 & 113 & 21132 & 54126 & 7\\
\hline
S5 & 91 & 186 & 181 & 114 & 2183 & 11582 & 2\\
\hline
S6 & 92 & 187 & 179 & 114 & 171 & 4329 & 1\\
\hline
\end{tabular}
\label{tab:healcare-smt-metrics}
\end{table}

 \subsection{Threats to Validity}

The complexity of designing and implementing goal and feature models introduces potential threats to internal validity. Our framework relies on expert judgment to model contextual factors, requirements, security risks, authentication methods, and their interdependencies. This reliance may introduce inaccuracies in their representation, potentially affecting the validity of the selected authentication methods. Ensuring internal validity, therefore, depends on the accurate modelling and interpretation of these elements. To mitigate this threat, in future work, we will focus on empirically estimating the impact of authentication methods on both security risk mitigation and requirement satisfaction. We will investigate data-driven approaches to calibrate the relationships between contextual factors, goal priorities, and authentication methods. Possible directions include leveraging historical authentication logs, user behavior analytics, and supervised learning techniques to infer the impact of context on goal satisfaction and method feasibility. In particular, for the Internet of Vehicles (IoV) domain, we plan to conduct empirical experiments using simulation platforms such as CARLA~\cite{dosovitskiy2017carla}(autonomous vehicle simulation), Veins~\cite{sommer2011bidirectionally} (vehicular network simulation combining OMNeT++~\cite{varga2010omnet++} and SUMO~\cite{lammel2017simulation}) or NS-3~\cite{riley2010ns}(network simulation). These platforms will enable us to simulate dynamic contextual changes and launch controlled IoV attacks, thereby evaluating how selected authentication methods mitigate security risks under varying conditions. In addition, we will analyze authentication logs—such as device type, network location, time of access, and authentication outcomes—to estimate the feasibility of authentication methods and their impact on performance requirements across different contexts. We will also empirically measure performance and usability satisfaction, including through user studies using established instruments such as the System Usability Scale (SUS).

Our evaluation currently assesses the effectiveness of authentication method selection based on the utility values produced by the proposed model. While this demonstrates that the approach aligns with the specified risks and goal priorities, it does not constitute validation against an external ground truth, such as expert assessments or established security best practices, thereby posing a threat to construct validity. As a result, “effectiveness” in our study reflects alignment with the model assumptions rather than an objectively optimal authentication choice. Future work will address this limitation by incorporating expert-based validation and comparisons with reference security guidelines. Another threat to construct validity arises from the correctness of the Z3 encoding used to enable FCN reasoning. Accurate translation of goal and feature models into Z3 constraints is essential for reliable results. To reduce this risk, two authors independently reviewed and validated the Z3 models to ensure correctness and consistency.

To mitigate threats to external validity, we applied our framework to two distinct domains (IoV and healthcare domains). Although additional domains (e.g., smart buildings) could be considered, demonstrating applicability across two different domains strengthens the generalizability of our approach. Nevertheless, reliance on expert judgment may still limit scalability and generalization. To address this threat, future work will investigate transfer learning techniques to reuse FCN representations across use cases. We also plan to refine impact weights using reinforcement learning informed by empirical data on attack success rates and requirement satisfaction under different contextual scenarios. Finally, we acknowledge that the performance of the Z3 solver may degrade for large constraint systems, posing an additional threat to external validity. To mitigate this issue, we will explore parallelization techniques, which have been shown to improve the performance of Z3 models addressing linear constraints with real variables~\cite{Delov.2025}. To mitigate this threat to validity, we will also explore the use of alternative approaches, such as an optimization-based solver to find a maximum utility value.

\section{Related Work}

Researchers have developed numerous adaptive authentication systems to address the trade-off between authentication needs (such as security and usability).  The prior work on adaptive authentication was focused on assessing contextual elements to evaluate the related risk and then selecting appropriate authentication mechanisms. Bakar and Haron \cite{bakar2014adaptive} used historical user behavior records (login time and location) to calculate a trust score and decide whether to use the same authentication method or change it based on the current context. Haron et al. \cite{daud2017adaptive} enhanced the work of \cite{bakar2014adaptive} by introducing multiple authentication techniques and single sign-on functionalities. Haron et al. \cite{daud2019adaptive} proposed a trust engine that only permits trusted IP addresses to login. Gebrie and Abie \cite{gebrie2017risk} analyzed user activity to assess the risk of log-in attempts, data loss, hacking, impersonation, eavesdropping, data extraction, and patient endangerment. The model chooses an appropriate authentication strategy depending on the risk score received.

Some researchers considered the user context-aware(e.g., location) to build the adaptive authentication. For example, Hulsebosch et al.~\cite{hulsebosch2007context} presented a system that enhances the effectiveness of traditional security measures by adapting its authentication process to a user's location through the use of identity tokens. By utilizing context information, the system makes these measures more adaptable to the current situation and less intrusive. The system strengthens its confidence in the user's identity by combining location information from various sources associated with the user. The likelihood of the user being in a specific location is used to determine the user's authentication level, allowing for more adaptive authentication that can adjust to changing circumstances. Lenzini et al.~\cite{lenzini2008trust} presented an adaptive authentication technique employed trust as a means of enhancing security during the authentication process. The method evaluates the user's identity by considering detectable accredited items and personal devices (e.g., (i.e., PDA, badge, laptop, Bluetooth devices, RFID) that the user carries, as well as position information. This approach enables versatile and adaptable control procedures that can adjust to changing circumstances. Seifert et al. \cite{seifert2010treasurephone} introduced an adaptive authentication approach that splits the user's data into spheres depending on the data connected to home and work and allows users to access these spheres based on the location. Das et al. \cite{hayashi2013casa} identified smartphone users using three geographic contexts: home, work, and other. The locking screen toggles between (no authentication, PIN, and password). 

To improve the user experience, Qin et al. \cite{riva2012progressive} proposed a progressive authentication technique that minimizes the number of authentication requests and makes authentication more comfortable for users who do not presently utilize any security mechanism on their devices. The method used numerous authentication signals, such as biometrics, continuity, and ownership, to assess a level of trust in a user's validity. The system assesses whether application access requires authentication depending on the application's confidence level and the level of security chosen. Valero et al. \cite{jorquera2018improving} proposed an adaptive continuous system that allows the mobile device to identify the owner periodically using machine learning techniques based on the anomaly detection approach. The system calculates the authentication level score using the user behaviour of application usage statistics that have been collected to recognize the user behaviour, whether it is new or old. Then, the system can adapt to the new behaviour as new behaviour or reject it.  Wójtowicz and Joachimiak \cite{wojtowicz2016model} proposed adaptive authentication to adjust the smartphone lock mechanism between voice recognition, face scan, and fingerprint, depending on which one is more usable for the current context based on sensed activity, light level, noise level, etc.  However, none of these works considered representing the different contextual factors, requirements, and authentication methods and their mutual impacts.

However, none of these works on adaptive authentication systems has provided extensive guidance on several key aspects. Firstly, there is a need to understand how to represent the various contextual factors, requirements, authentication methods, and the relationships between them. Secondly, it is crucial to identify how these factors affect the fulfillment of adaptive authentication requirements and how these impacts would be weighted and represented. Lastly, it is necessary to determine the optimal authentication method that maximizes the satisfaction of these requirements. While also previous work on adaptive systems, such as~\cite{bucchiarone2010context,tamura2013improving,kulp2021towards,Salehie.RE.2012,pasquale2016automating}, have primarily focused on context-driven adaptation, they have not fully considered the impact of context on the relative importance of requirements and the feasibility of authentication methods. Additionally, the highly personal nature of authentication means that users' preferences and security requirements can significantly influence adaptation decisions.

\section{Conclusion}

In this article, we proposed a framework for adaptive authentication to select the most appropriate authentication method to mitigate security risks and optimize the satisfaction of security, usability and performance goals. We use a contextual goal model to represent the system goals and the impact contextual factors have on the goals' priorities. Additionally, we used an extended feature model to describe the features characterizing the authentication methods that can be selected. We extend this feature model with an explicit representation of the impact of contextual factors on the enabling/disabling of specific authentication methods. We used a Fuzzy Causal Network (FCN) encoded with Z3, an efficient SMT solver, to compute the total risk and select an effective authentication method that maximises utility, representing the trade-off between satisfying the system goals and reducing security risk. We assumed that contextual factors are monitored at runtime, and that whenever they change, the FCN analysis is triggered to select an authentication method.
We evaluated our approach using two case studies considering the IoV and the healthcare domains. We compared our approach with a baseline that selects the same authentication method for each scenario considered in the case study. We demonstrate that our approach outperforms the baseline at reducing security risks while ensuring the highest-priority system goals are met. It also incurs a small time and memory overhead, demonstrating feasibility for runtime deployment across the considered case studies.


\section{Acknowledgement}

This publication was supported partially by the EU H2020 CyberSec4Europe project (grant number $830929$) and by Taighde {E}ireann $–$ Research Ireland under Grant number $13/$RC$/$$2094_2$. 
\bibliographystyle{ACM-Reference-Format}
\bibliography{Ref}

\end{document}